\documentclass[acmsmall]{acmart}
\AtBeginDocument{%
  }

\setcopyright{none}
\copyrightyear{XXXX}
\acmYear{XXXX}
\acmDOI{XXXXXXX.XXXXXXX}

\acmJournal{TOSEM}
\acmVolume{X}
\acmNumber{X}
\acmArticle{X}
\acmMonth{0}

\usepackage{braket}
\usepackage{longtable}
\usepackage{soul}

\def\h{\ensuremath{\mathcal{H}}}
\def\l{\ensuremath{\mathcal{L}}}
\def\lh{\ensuremath{\mathcal{L(H)}}}
\newcommand{\tr}{{\rm tr}}
\newcommand{\free}{{\rm free}}
\newcommand{\modi}{{\rm mod}}
\def\Qimp{{\textbf{QIMP}}}
\def\Cvar{{\textbf{Cvar}}}
\def\Qvar{{\textbf{Qvar}}}
\def\Aexp{{\textbf{Aexp}}}
\def\Bexp{{\textbf{Bexp}}}
\def\Com{{\textbf{Com}}}
\def\Skip{{\textbf{skip}}}

\def\abort{\textbf{abort}}
\def\While{\textbf{while}~}
\def\Do{~\textbf{do}~}
\def\If{\textbf{if}~}
\def\Then{~\textbf{then}~}
\def\Else{~\textbf{else}~}
\def\Od{~\textbf{od}~}

\def\CNOT{{\textit{CNOT}}}
\renewcommand{\>}{\ensuremath{\rangle}}
\newcommand{\<}{\ensuremath{\langle}}
\def\pair#1{\langle#1\rangle}
\def\true{{\textbf{true}}}
\def\false{{\textbf{false}}}
\newcommand{\tU}{{ U}}
\newcommand{\tM}{{ M}}
\def\CH{{\mathcal H}}
\def\qstate{\rho}
\def\state{\sigma}
\def\PState{{\textbf{POVD}}}
\def\povd{\mu}
\def\Nat{{\mathbb N}}
\newcommand{\padist}[1]{\mathop{\mbox{$\mathcal D^-$}} ({#1})} 
\newcommand{\size}[1]{\| #1\|}
\newcommand{\support}[1]{\lceil{#1}\rceil}
\newcommand{\denote}[1]{[\![#1]\!]}
\newcommand{\Key}[1]{{\color{blue} #1}}
\newcommand{\In}{{\color{brown} \Longrightarrow}}
\newcommand{\Out}{{\color{brown} \Longleftarrow}}
\newcommand{\InOut}{{\color{brown} \Longleftrightarrow}}

\newcommand{\leaveout}[1]{}
\def\sset#1{\{#1\}}

\begin{document}

\title[Local Reasoning for Classical--Quantum Programs]{Local Reasoning about Probabilistic Behaviour for Classical--Quantum Programs}
\titlenote{This work is an extended version of a paper that has appeared in the proceedings of the 25th International Conference on Verification, Model Checking, and Abstract Interpretation (VMCAI 2024).\\
Corresponding author: Yuxin Deng
}


\author{Huiling Wu}
\affiliation{%
  \institution{Shanghai Key Laboratory of Trustworthy Computing, East China Normal University}
  \city{Shanghai}
  \country{China}}
  \email{52275902009@stu.ecnu.edu.cn}

\author{Yuxin Deng}
\affiliation{%
  \institution{Shanghai Key Laboratory of Trustworthy Computing, East China Normal University}
  \city{Shanghai}
  % \state{Shanghai}
  \country{China}}
  \email{yxdeng@sei.ecnu.edu.cn}

\author{Ming Xu}
\affiliation{%
  \institution{Shanghai Key Laboratory of Trustworthy Computing, East China Normal University}
  \city{Shanghai}
  % \state{Shanghai}
  \country{China}}
  \email{mxu@cs.ecnu.edu.cn}

\renewcommand{\shortauthors}{H. Wu, Y. Deng, M. Xu}

\begin{abstract}
 Verifying the functional correctness of programs with both classical and quantum constructs is a challenging task. The presence of probabilistic behaviour entailed by quantum measurements and unbounded while loops complicate the verification task greatly. We propose a new quantum Hoare logic for local reasoning about probabilistic behaviour by introducing distribution formulas to specify probabilistic properties. We show that the proof rules in the logic are sound with respect to a denotational semantics. To demonstrate the effectiveness of the logic, we formally verify the correctness of non-trivial quantum algorithms including the HHL and Shor's algorithms. %Finally, 
 Moreover, we embed our logic into the proof assistant Coq. The resulting logical framework, called CoqQLR, can facilitate semi-automated reasoning about classical--quantum programs.
\end{abstract}


%%



\received{XXX}
\received[revised]{XXX}
\received[accepted]{XXX}

\maketitle

\section{Introduction}
Programming is an error-prone activity, and the situation is even worse for quantum programming, which is far less intuitive than classical computing. Therefore, developing verification and analysis techniques to ensure the correctness of quantum programs is an even more important task than that for classical programs.

Hoare logic~\cite{Hoa69} is probably the most widely used program logic to verify the correctness of programs. It is useful for reasoning about  deterministic and probabilistic programs. A lot of efforts have been made to reuse the nice idea to verify quantum programs.
Ying \cite{Yin12,Yin16} was the first to establish a sound and relatively complete quantum Hoare logic to reason about pure quantum programs, i.e., quantum programs without classical variables. This work triggered a series of research in this direction. For example, Zhou et al.~\cite{zhou2019applied} proposed an applied quantum Hoare logic by only using projections as preconditions and postconditions, which makes the practical use of quantum Hoare logic much easier. Barthe et al. \cite{Barthe2020}
extended quantum Hoare logic to relational verification by introducing a quantum analogue of probabilistic couplings.  Li and Unruh~\cite{LU21} defined a quantum relational Hoare logic with expectations in pre- and postconditions. Formalization of quantum Hoare logic in proof assistants such as Isabelle/HOL~\cite{NWP02} and Coq~\cite{BC04} was accomplished in \cite{Liu19} and \cite{Zhou23}, respectively.
Ying et al.~\cite{YZLF22} defined a class of disjoint parallel quantum programs and generalised the logic in \cite{Yin12} to this setting. Zhou et al.~\cite{Zhou21} developed a quantum separation logic for local reasoning.
%They and proved the relative completeness of the logic. 
However, all the work mentioned above cannot deal with programs that have both classical and quantum data.

In order to verify quantum programs found in almost all practical quantum programming frameworks such as Qiskit\footnote{\url{https://qiskit.org}}, $Q^\#$\footnote{\url{https://github.com/microsoft/qsharp-language}}, Cirq\footnote{\url{https://github.com/quantumlib/Cirq}}, etc., we have to explicitly consider programs with both classical and quantum constructs. Verification techniques for this kind of hybrid programs have been put forward in the literature \cite{CHADHA200619,Kaku09,Unr19,Unr2019,FY21,DENG202273,FLY22,Sun22}.
For example, Chadha et al.~\cite{CHADHA200619} proposed an ensemble exogenous quantum propositional logic  for a simple quantum language with bounded iteration. The expressiveness of the language is very limited, and algorithms involving unbounded while loops such as the HHL and Shor's algorithms~\cite{HHL09,Sho94} cannot be described. Kakutani~\cite{Kaku09} presented a quantum Hoare logic for an imperative language with while loops, but  the rule for them has no invariance condition. Instead, an infinite sequence of assertions has to be used. 
Unruh introduced a quantum Hoare logic with ghost variables to express properties such as that a quantum variable is disentangled with others \cite{Unr19} and a relational Hoare logic \cite{Unr2019} for security analysis of post-quantum cryptography and quantum protocols.
Deng and Feng~\cite{DENG202273} provided an abstract and a concrete proof system for classical--quantum programs, with the former being sound and relatively complete, while the latter being sound.
Feng and Ying~\cite{FY21} introduced classical--quantum assertions, which are a class of mappings from classical states to quantum predicates, to analyse both classical and quantum properties. The approach was extended to verify distributed quantum programs in \cite{FLY22}.
However, except for \cite{Sun22}, all the work above offers no support for local reasoning, which is an obvious drawback. In the case that we have a large quantum register but only a few qubits are modified, it is awkward to always reason about the global states of the quantum register. Based on this observation, Le et al. \cite{Sun22} provided an interesting quantum interpretation of the separating conjunction, so to infuse separation logic into a Hoare-style framework and thus support local reasoning. However, a weakness of their approach is that it cannot handle probabilistic behaviour, which exists inherently in quantum programs, in a satisfactory way. Let us illustrate this with a simple example.
\begin{example}\label{ex1}
The program \textbf{addM} defined below first initialises two qubits $q_0$ and $q_1$, and then applies the Hadamard gate $H$ to each of them. By measuring them with the measurement operators $|0\>\<0|$ and $|1\>\<1|$, we add the measurement results and assign the sum to the variable $v$. 
	\[
	\begin{aligned}
		\mathbf{addM} \ \triangleq \quad   
		&q_{0}:=\ket{0};\ H[q_0];\\
		&q_{1}:=\ket{0};\ H[q_1];\\
		&v_{0}:=M[q_{0}];\\
		&v_{1}:=M[q_{1}];\\
		&v:=v_{0}+v_{1}\\
	\end{aligned}
	\]
        Since the classical variable $v_0$ takes either $0$ or $1$ with equal chance, and similarly for $v_1$, the probability that variable $v$ is assigned to $1$ should be exactly $\frac{1}{2}$. However, in the program logic in \cite{Sun22}, this property cannot be specified.
\end{example}

We propose a novel quantum Hoare logic for a classical--quantum language.
Two distribution formulas $\oplus_{i\in I}p_i\cdot F_i$ and $\oplus_{i\in I} F_i$ are introduced. A program state $\mu$, which is a partial density operator valued distribution (POVD) \cite{DENG202273}, satisfies the formula $\oplus_{i\in I}p_i\cdot F_i$ if $\mu$ can be split into the weighted sum of %$i$ parts called 
some $\mu_i$ with weights $p_i$.
Each $\mu_i$ has the same size as $\mu$ and satisfies the formula $F_i$ whenever $p_i > 0$. A state $\mu$ satisfies the formula $\oplus_{i\in I} F_i$ if there exists a collection of probabilities $\{p_i\}_{i\in I}$ with $\sum_{i\in I}p_i= 1$ such that  $\oplus_{i\in I}p_i\cdot F_i$ can be satisfied. In other words, the splitting of $\mu$ does not necessarily follow a fixed set of weights. With distribution formulas, we can conveniently reason about the probabilistic behaviour mentioned in Example~\ref{ex1} (more details will be discussed in Example~\ref{ex3}), and give an invariance condition in the proof rule for while loops. In addition, we adopt the labelled Dirac notation emphasised in \cite{Zhou23} to facilitate local reasoning. Our program logic is shown to be sound and can be used to prove the correctness of non-trivial quantum algorithms including the HHL and Shor's algorithms. Furthermore, recognizing the complexity and laboriousness of manual reasoning about these algorithms, we %integrate 
embed
our logic into the Coq proof assistant. The resulting logical framework, called CoqQLR, can facilitate semi-automated reasoning. Each rule in our proof system is formulated as a theorem and rigorously validated for soundness within the Coq environment. Ultimately, we formally demonstrate the correctness of the aforementioned algorithms with machine-checkable proofs in Coq.

Therefore, the main contributions of the current work include the following aspects:
\begin{itemize}
\item We propose to use distribution formulas in a new quantum Hoare logic to specify the probabilistic behaviour of classical--quantum programs. Distribution formulas are useful to give an invariance condition in the proof rule for while loops, so to avoid an infinite sequence of assertions in the rule.
\item We prove the soundness of our logic that enables local reasoning in the spirit of separation logic. A crucial element of the proof is the property that the denotation of each assertion constitutes a closed set.
  \item We demonstrate the effectiveness of the logic by proving the correctness of the HHL and Shor's algorithms.
  \item We integrate our logic into the proof assistant Coq, enabling semi-automated reasoning about classical--quantum programs.
\end{itemize}

The rest of the paper is structured as follows. In Section~\ref{sec:pre} we recall some basic notations about quantum computing. In Section~\ref{sec:lang} we review the syntax and denotational semantics of a classical--quantum imperative language considered in~\cite{DENG202273}.
In Section~\ref{sec:proof} we define an assertion language and propose a proof system for local reasoning about quantum programs.
We also prove the soundness of the system. In Section~\ref{sec:cases} we apply our program logic to verify the HHL and Shor's algorithms.  
Section~\ref{sec:imple} is dedicated to the Coq formalization of our logic. 
Finally, we conclude in Section~\ref{sec:con} and discuss possible future work. Our Coq development is available at the following link:\\
\centerline{\url{https://github.com/fox9909/CoqQLR.git}}

% \yx{[It seems that the link doesn't work!]}

\section{Preliminaries}\label{sec:pre}
We briefly recall some basic notations
from linear algebra and quantum mechanics which are needed in this paper. 
For more details, we refer to \cite{NC00}.

A {\it Hilbert space} $\h$ is a complete vector space with an inner
product $\langle\cdot|\cdot\rangle:\h\times \h\rightarrow \mathbb{C}$
such that 
\begin{enumerate}
\item
$\langle\psi|\psi\rangle\geq 0$ for any $|\psi\>\in\h$, with
equality if and only if $|\psi\rangle =0$,
\item
$\langle\phi|\psi\rangle=\langle\psi|\phi\rangle^{\ast}$,
\item
$\langle\phi|\sum_i c_i|\psi_i\rangle=
\sum_i c_i\langle\phi|\psi_i\rangle$,
\end{enumerate}
where $\mathbb{C}$ is the set of complex numbers, and for each
$c\in \mathbb{C}$, $c^{\ast}$ stands for the complex
conjugate of $c$. For any vector $|\psi\rangle\in\h$, its
length $\||\psi\rangle\|$ is defined to be
$\sqrt{\langle\psi|\psi\rangle}$, and it is said to be {\it normalised} if
$\||\psi\rangle\|=1$. Two vectors $|\psi\>$ and $|\phi\>$ are
{\it orthogonal} if $\<\psi|\phi\>=0$.
An {\it orthonormal basis} of a Hilbert
space $\h$ is a basis $\{|i\rangle\}$ where each $|i\>$ is
normalised and any pair of them are orthogonal.

Let $\lh$ be the set of linear operators on $\h$.
For any $A\in \lh$, $A$ is {\it Hermitian} if $A^\dag=A$ where $A^\dag$ is the adjoint operator of $A$ such that
$\<\psi|A^\dag|\phi\>=\<\phi|A|\psi\>^*$ for any
$|\psi\>,|\phi\>\in\h$.
A linear operator $A\in \lh$ is {\it unitary} if $A^\dag A=A A^\dag=I_\h$ where $I_\h$ is the
identity operator on $\h$. 
The {\it  trace} of $A$ is defined as $\tr(A)=\sum_i \<i|A|i\>$ for some
given orthonormal basis $\{|i\>\}$ of $\h$.
A linear operator $A\in \lh$ is {\it positive} if $\<\phi|A|\phi\> \geq 0$ for any vector $|\phi\> \in\h$.
The \emph{L\"{o}wner order} $\sqsubseteq$ on the set of Hermitian operators on $\h$ is defined
by letting $A\sqsubseteq B$ if and only if $B-A$ is positive.

Let $\h_1$ and $\h_2$ be two Hilbert spaces. Their {\it tensor product} $\h_1\otimes \h_2$ is
defined as a vector space consisting of
linear combinations of the vectors
$|\psi_1\psi_2\rangle=|\psi_1\>|\psi_2\rangle =|\psi_1\>\otimes
|\psi_2\>$ with $|\psi_1\rangle\in \h_1$ and $|\psi_2\rangle\in
\h_2$. Here the tensor product of two vectors is defined by a new
vector such that
\[\left(\sum_i \lambda_i |\psi_i\>\right)\otimes
\left(\sum_j\mu_j|\phi_j\>\right)=\sum_{i,j} \lambda_i\mu_j
|\psi_i\>\otimes |\phi_j\>.\] Then $\h_1\otimes \h_2$ is also a
Hilbert space where the inner product is defined as the following:
for any $|\psi_1\>,|\phi_1\>\in\h_1$ and $|\psi_2\>,|\phi_2\>\in
\h_2$,
\[\<\psi_1\otimes \psi_2|\phi_1\otimes\phi_2\>=\<\psi_1|\phi_1\>_{\h_1}\<
\psi_2|\phi_2\>_{\h_2}\] where $\<\cdot|\cdot\>_{\h_i}$ is the inner
product of $\h_i$. Given $\h_1$ and $\h_2$, the \emph{partial trace} with respect to $\h_2$, written $\tr_{\h_2}$, is a linear mapping from $\l(\h_1\otimes \h_2)$ to $\l(\h_1)$ such that for any $|\psi_1\>,|\phi_1\>\in\h_1$ and $|\psi_2\>,|\phi_2\>\in
\h_2$,
\[\tr_{\h_2}(|\psi_1\>\<\phi_1|\otimes |\psi_2\>\<\phi_2|) = \<\psi_2|\phi_2\>|\psi_1\>\<\phi_1|.\]

By applying quantum gates to qubits, we can change their
states. For example, the Hadamard gate ($H$ gate) can be applied
on a single qubit, while the controlled-NOT gate ($\CNOT$ gate) can be applied on two
qubits. Their representations in terms of matrices are given as
\[
H=\tfrac{1}{\sqrt{2}}
\begin{pmatrix}
  1 & 1 \\
  1 & -1 \\
\end{pmatrix} \quad \textup{and} \quad
\CNOT=\begin{pmatrix}
  1 & 0 & 0 & 0 \\
  0 & 1 & 0 & 0 \\
  0 & 0 & 0 & 1 \\
  0 & 0 & 1 & 0
\end{pmatrix}.
\]

\leaveout{%1307
\[
H=\frac{1}{\sqrt{2}}\left(%
\begin{array}{cc}
  1 & 1 \\
  1 & -1 \\
\end{array}%
\right),\ \  I_2=\left(%
\begin{array}{cc}
  1 & 0 \\
  0 & 1 \\
\end{array}%
\right),\ \
X=\left(%
\begin{array}{cc}
  0 & 1 \\
  1 & 0 \\
\end{array}%
\right),\ \ Z=\left(%
\begin{array}{cc}
  1 & 0 \\
  0 & -1 \\
\end{array}%
\right).
%,\ Y=\left(%
%\begin{array}{cc}
%  0 & -i \\
%  i & 0 \\
%\end{array}%
%\right).
\]
} %endofleaveout 1307

According to von Neumann's formalism of quantum mechanics
\cite{vN55}, an isolated physical system is associated with a
Hilbert space which is called the {\it state space} of the system. A {\it pure state} of a
quantum system is a normalised vector in its state space, and a
{\it mixed state} is represented by a density operator on the state
space. Here a {\it density operator} $\rho$ on Hilbert space $\h$ is a
positive linear operator such that $\tr(\rho)= 1$. A {\it partial density operator} $\rho$ is a positive linear operator with $\tr(\rho)\leq 1$.

The evolution of a closed quantum system is described by a unitary
operator on its state space: if the states of the system at times
$t_1$ and $t_2$ are $\rho_1$ and $\rho_2$, respectively, then
$\rho_2=U\rho_1U^{\dag}$ for some unitary operator $U$ which
depends only on $t_1$ and $t_2$. 
%In contrast, the general dynamics which can occur in a physical system is described by a trace-preserving super-operator on its state space. Note that the unitary transformation $U(\rho)=U\rho U^\dag$ is a trace-preserving super-operator. 
A quantum {\it measurement} is described by a
collection $\{M_m\}$ of positive operators, called measurement operators,
where the indices $m$ refer to the measurement outcomes. It is required that the
measurement operators satisfy the completeness equation
$\sum_{m}M_m^{\dag}M_m=I_\h$. If the system is in state $\rho$, then the probability
that measurement result $m$ occurs is given by
\[p(m)=\tr(M_m^{\dag}M_m\rho),\] and the state of the post-measurement system
is $M_m\rho M_m^{\dag}/p(m).$ 

\section{A Classical--Quantum Language}\label{sec:lang}
Here, we recall the simple classical--quantum imperative language  {\Qimp} as defined in \cite{DENG202273}.
It is essentially similar to a few imperative languages considered in the literature \cite{Sel04,YF11,Unr2019,FY21}.
We introduce its syntax and denotational semantics.

\subsection{Syntax}

\begin{table}[t]
     \centering
     \caption{Syntax of quantum programs}
     \label{tab:syn_pro}
    \[
	\begin{aligned}
		&(\Aexp)& \ &a& \ &::=& \ & n \ |\ x,y,...\ |\ f_{m}(a,...,a)\\
		&(\Bexp)& \ &b&  \
		&::=& \ &\textbf{true}\ |\ \textbf{false}\ |\ P_{m}(a,...,a)\ |\ b \wedge b \ | \ \neg b \ | \ \forall x.b\\
		&(\Com)& \ &c& \ &::=& \ &\textbf{skip} \ |\ \mathbf{abort} \ |\ x:=a \ |\ c ;c \\
		&&&&&&& | \  \textbf{if} \ b \ \textbf{then} \ c \ \textbf{else} \ c \  \mathbf{fi} \ |\ \textbf{while} \ b \  \textbf{do} \ c \ \mathbf{od}\\
		&& && && &|\ q :=\ket{0} \ |\ U[\overline{q}]\ |\ x:=M[\overline{q}]
	\end{aligned}
	\]
 \end{table}

We assume three types of data in our language: {\tt Bool} for booleans,  {\tt Int} for integers, and  {\tt Qbt} for quantum bits (qubits). Let $\mathbb{Z}$ be the set of  integer constants, ranged over by $n$. Let $\Cvar$, ranged over by $x,y,...$, be the set of classical variables, and $\Qvar$, ranged over by $q,q',...$, the set of quantum variables. It is assumed that both $\Cvar$ and $\Qvar$ are countable. We assume a set {\Aexp} of arithmetic expressions over {\tt Int}, which includes $\Cvar$ as a subset and is ranged over by $a, a',...$, and a set of boolean-valued expressions $\Bexp$, ranged over by $b,b',...$, with the usual boolean constants {\true, \false} and boolean connectives such as $\neg, \wedge$ and $\vee$. 
We assume a set of arithmetic functions (e.g. $+$, $-$, $*$, etc.) ranged over by the symbol $f_m$,
and a  set of boolean predicates (e.g. $=$, $\leq$, $\geq$, etc.) ranged over by $P_m$, where $m$ indicates the number of variables involved.
We further assume that only classical variables can occur freely in both arithmetic and boolean expressions. 

We let $U$ range over unitary operators, which can be  user-defined  matrices or built in if the language is implemented. For example, a concrete $U$ could be the $1$-qubit Hadamard operator $H$, or the 2-qubit controlled-NOT operator $\CNOT$, etc. Similarly, we write $M$ for the measurement described by a collection $\{M_i\}$ of measurement operators, with each index $i$ representing a measurement outcome. For example, to describe the measurement of the qubit referred to by variable $q$ in the computational basis, we can write $M:=\{M_0, M_1\}$, where $M_0=|0\>_q\<0|$ and $M_1=|1\>_q\<1|$.

Sometimes, we use metavariables which are primed or subscripted, e.g. $x', x_0$ for classical variables.
We abbreviate a tuple of quantum variables $\pair{q_1,...,q_n}$ as $\bar{q}$ if the length $n$ of the tuple is not important. If two tuples of quantum variables $\overline{q}$ and $\overline{q^{'}}$  are disjoint, where $\overline{q}$=$\langle q_{1},...,q_{n}\rangle $ and $\overline{q^{'}}=\langle q_{n+1},q_{n+2},...,q_{n+m}\rangle $, then their concatenation is a larger tuple $\overline{q}$$\overline{q^{'}}$=$\langle q_{1},...,q_{n},q_{n+1},...,q_{n+m}\rangle $. If no confusion arises, we occasionally use a tuple to stand for a  set.

The formation rules for arithmetic and boolean expressions as well as commands are defined in Table~\ref{tab:syn_pro}.
An arithmetic expression can be an integer, a variable, or built from other arithmetic expressions  by some arithmetic functions.  A boolean expression can be a boolean constant, built from arithmetic expressions by some boolean predicates or formed by using the usual boolean operations. A command can be a skip statement, an abort statement, a classical assignment, a conditional statement, or  a while-loop, as in many classical imperative languages. The command $\mathbf{abort}$ represents the unsuccessful termination of programs. In addition, there are three commands that involve quantum data. The command $q := |0\>$ initialises the qubit referred to by variable $q$ to be the basis state $|0\>$. The command $\tU[\bar{q}]$ applies the unitary operator $\tU$ to the quantum system referred to by $\bar{q}$. The command $x := \tM[\bar{q}]$ performs a measurement $M$ on $\bar{q}$ and assigns the measurement outcome to $x$. It differs from a classical assignment because the measurement $M$ may change the quantum state of $\bar{q}$, besides the fact that the value of $x$ is updated.

  	For convenience, we further define the following syntactic sugar for the initialization of a sequence of quantum variables:
        $\overline{q}=\ket{0}^{\otimes n}$,
        where $\overline{q}=\langle q_{1},...,q_{n} \rangle$, is an abbreviation of the commands:
	\[
	q_{1}=\ket{0}; q_{2}=\ket{0}; \ldots; q_{n}:=\ket{0}.
	\]

 \subsection{Denotational Semantics}
In the presence of classical and quantum variables, the execution of a {\Qimp} program involves two types of states: classical states and quantum states.

As usual, a classical state is a function $\sigma: \Cvar\rightarrow\mathbb{Z}$ from classical variables to integers, where $\sigma(x)$ represents the value of classical variable $x$.
For each quantum variable $q\in \Qvar$, we assume a $2$-dimensional Hilbert space $\CH_q$ to be the state space of the
$q$-system. For any finite subset $V$ of $\Qvar$, we denote
\[ \CH_V \ = \ \bigotimes_{q\in V}\CH_q.\] 
That is, $\CH_V$ is the Hilbert space spanned by tensor products of the individual state spaces of the quantum variables in $V$. Throughout the paper, when we refer to a subset of $\Qvar$, it is assumed to be finite.
 Given $V\subseteq\Qvar$, the set of \emph{quantum states} consists of all partial density operators in the space $\CH_V$, denoted by $\padist{\CH_V}$.  For any quantum state $\rho \in \padist{\CH_V}$, its \textit{domain} is  defined as $\mathrm{dom}(\rho) \triangleq V$. Sometimes, we simply use the notation $\rho_{\overline{q}}$ to denote a quantum state $\rho$ such that $\rho \in \padist{\CH_{\overline{q}}}$. Similarly, we use  $
 \rho_{V}$ to denote some $\rho \in \padist{\CH_V}$.
 Let $V'$ be a set of quantum variables with $V'\subseteq V$. We write $\rho|_{V'}$ for the reduced density operator $\tr_{V\backslash V'}(\rho)$ obtained by restricting $\rho$ to $V'$.
 
 For any set $V \subseteq \mathbf{QVar}$, a \emph{machine state} in the space  $\CH_V$ is a pair $\pair{\state,\qstate}$, where $\state$ is a classical state and $\qstate$ is a quantum state in the space $\CH_V$. Similarly, for any state $s$ in the space  $\CH_V$, its \textit{domain} is defined as $\mathrm{dom}(s)\triangleq V$.  In the presence of measurements, we often need to consider an ensemble of states. For that purpose, we introduce a notion of distribution.
 \begin{definition}\cite{DENG202273}
 Suppose $V\subseteq\Qvar$ and $\Sigma$ is the set of classical states, i.e., the set of functions of type $\Cvar\rightarrow\mathbb{Z}$.
 A \emph{partial density operator valued distribution in the space $\CH_V$\ (POVD$_V$)} is a function $\povd: \Sigma\rightarrow\padist{\CH_V}$ with $\sum_{\sigma\in\Sigma}\tr(\mu(\sigma))\leq 1$.
 \end{definition} 
 Intuitively, a POVD$_V$ $\povd$ represents a collection of machine states with domain $V$, where each classical state $\sigma$ is associated with a quantum state $\povd(\sigma)$. We write $\PState_V$ for the set of POVD$_V$. The subscript $V$ is utilized to indicate the specific space. When the context makes it clear or the space is not important, we may simply refer to it as POVD. 
This notation of POVD is called  classical--quantum state in \cite{FY21}. In this paper, we sometimes call it a distribution state and abbreviate it as distribution.
 If the collection has only one element $\sigma$, we explicitly write $(\sigma,\povd(\sigma))$ for $\povd$. The support of $\povd$, written $\support{\povd}$, is the set $\{\sigma\in\Sigma \mid \povd(\sigma)\neq 0\}$. The size of $\povd$, written $\size{\povd}$, means $\sum_{\sigma\in\Sigma}\tr(\povd(\sigma))$.
 We write $\varepsilon$ for the special POVD whose support is the empty set. Similarly, for any distribution $\mu$ in the space $\CH_V$, its \textit{domain} is defined as $\mathrm{dom}(\mu)\triangleq V$. The notation $\mu_V$ is used to denote the distribution $\mu$ with domain $V$. Let $V'$ be a subset of quantum variables such that $V'\subseteq V$. We denote the reduced distribution of $\mu$ on $V'$ by $\mu|_{V'}$, which is a POVD$_{V'}$ defined by the equation $\mu|_{V'}(\sigma)= (\mu(\sigma))|_{V'}$. We can also define the scaling of a distribution by letting $(p \cdot \mu) (\sigma) = p \cdot (\mu (\sigma )) $ for some probability $p$ and the addition of two distributions by letting
 $(\povd_1+\povd_2)(\sigma) = \povd_1(\sigma)+\povd_2(\sigma)$.

\begin{table}[t]
\caption{Denotational semantics of commands}
\label{tab:den}
\[\begin{array}{rcl}
\denote{\Skip}_{(\state,\qstate)} & = & (\state,\qstate)  \\
\denote{\abort}_{(\state,\qstate)}  & = & \varepsilon \\
\denote{x := a}_{(\state,\qstate)}  & = & (\state[\denote{a}_\state/x], \qstate) \\
\denote{c_0;c_1}_{(\state,\qstate)}  & = & \denote{c_1}_{\denote{c_0}_{(\state,\qstate)} }\\
\denote{\If b \Then c_0 \Else c_1\ {\bf fi}}_{(\state,\qstate)}  & = & \left\{\begin{array}{ll}
\denote{c_0}_{(\state,\qstate)}  & \mbox{if $\denote{b}_\state=\true$}\\
 \denote{c_1}_{(\state,\qstate)}  & \mbox{if $\denote{b}_\state=\false$}
        \end{array}\right.\\
\denote{\While b \Do c\ {\bf od}}_{(\state,\qstate)}  & = &  \lim_{n\rightarrow\infty}
\denote{(\If b \Then c\ {\bf fi})^n; \If b \Then \abort\ {\bf fi}}_{(\state,\qstate)} \\
\denote{q := |0\>}_{(\state,\qstate)}  & = &  (\state,\qstate')\\
& & \mbox{where $\qstate' :=  |0\>_q\<0|\qstate |0\>_q\<0| + |0\>_q\<1|\qstate |1\>_q\<0| $} \\
\denote{\tU[\bar{q}] }_{(\state,\qstate)}  & = & (\state, \tU\qstate\tU^\dag)\\
\denote{x := \tM[\bar{q}]}_{(\state,\qstate)}  & = & \povd \\
& & \mbox{where $M=\sset{M_i}_{i\in I}$ and} \\
  & & \povd(\sigma')=\sum_i\{M_i \qstate M_i^\dag \mid \state[i/x]=\sigma'\}\\
\denote{c}_{\povd} & = & \sum_{\state\in\support{\povd}} \denote{c}_{(\state,\povd(\state))}.
\end{array}\]
\end{table}

We interpret a program $c$ as POVD transformers.
Given an expression $e$, we denote its interpretation with respect to machine state $(\state,\qstate)$ by $\denote{e}_{(\state,\qstate)}$.  For any program $c$, we denote the set of quantum variables modified by $c$ as qmod($c$) and the set of classical and quantum variables modified by $c$ as mod($c$).
Let $V$ be a set of quantum variables. For any program $c$ with $\mathrm{qmod}(c) \subseteq V$, we define the denotational semantics of $c$ with domain $V$ as a mapping $[\![c]\!]_{V}: \PState_V \rightarrow \PState_V$, displayed in Table~\ref{tab:den}. Sometimes, we omit the subscript $V$ when it can be inferred from the context. Additionally, here we omit the denotational semantics of arithmetic and boolean expressions such as $\denote{a}_\state$ and $\denote{b}_\state$, which is almost the same as in the classical setting because the quantum part  plays no role in those expressions. A state evolves into a POVD after some quantum qubits are measured, with the measurement outcomes assigned to a classical variable. Two other quantum commands, initialisation of qubits and unitary operations, are deterministic and only affect the quantum part of a state. As usual, we define the semantics of a loop ($\While b \Do c\ {\bf od}$) as the limit of its lower approximations, where the $n$-th lower approximation of $\denote{\While b \Do c\ {\bf od}}_{(\state,\qstate)} $ is $\denote{(\If b \Then c\ {\bf fi})^n; \If b \Then \abort\ {\bf fi}}_{(\state,\qstate)} $, where ($\If b \Then c\ {\bf fi}$) is shorthand for ($\If b \Then c \Else \Skip\ {\bf fi}$) and $c^n$ is the command $c$ iterated $n$ times with $c^0\equiv\Skip$. The limit exists because the sequence $(\denote{(\If b \Then c\ {\bf fi})^n; \If b \Then \abort\ {\bf fi}}_{(\state,\qstate)} )_{n\in\Nat}$ is strictly increasing and bounded with respect to the L\"{o}wner order \cite[Lemma 3.2]{DENG202273}. 

We remark that the semantics $\denote{c}_{(\state,\qstate)} $ of a command $c$ in initial state ${(\state,\qstate)} $ is a POVD.
The lifted semantics $\denote{c}_\povd$ of a command $c$ in initial POVD $\povd$ is also a POVD.
Furthermore, the function $\denote{c}$ is linear in the sense that \[  [\![c]\!]_{\sum_i p_i \cdot \mu_i} =\sum_i p_i \cdot  [\![c]\!]_{\mu_i}. \]
% \change{where $p_i > 0$ for each $i$ and $\sum_i p_i \leq 1$.} \change{[Because it doesn't hold when some $p_i = 0 $. ]}
% where we write $p\mu$ for the POVD defined by $(p\mu)(\sigma)=p\cdot \mu(\sigma)$.
Additionally, the function \(\denote{c}\) commutes with the restriction of distributions, i.e.,
\[
[\![c]\!]_{(\mu|_{V})} = ([\![c]\!]_{\mu})|_{V},
\]
for any \(V\) such that \(\mathrm{qmod}(c) \subseteq V \subseteq \mathrm{dom}(\mu)\).

Similarly as in~\cite{Zhou23},
we take advantage of the labelled Dirac notation throughout the paper with subscripts identifying the subsystem where a
ket/bra/operator lies or operates. For example,
%	The Dirac notation $\ket{a}$ is any normalized vector in a Hilbert space. We use the Dirac notation with a 
the subscript in $\ket{a}_{\overline{q}}$ indicates the Hilbert space $\mathcal{H}_{\overline{q}}$ where the state $\ket{a}$ lies. The notation $\ket{a}\ket{a}$ and $\ket{a}_{\overline{q}}$$\ket{a}_{\overline{q}}$ are the abbreviations of $\ket{a}\otimes\ket{a}$ and  $\ket{a}_{\overline{q}}$$\otimes$$\ket{a}_{\overline{q}}$ respectively. We also use  operators with subscripts %such as $_{\overline{q}}\bra{a}$ $\ket{a}_{\overline{q}}\bra{a}$, $\langle {a}|{a} \rangle_{\overline{q}}$ and
like $A_{\overline{q}}$ to identify the Hilbert space  $\mathcal{H}_{\overline{q}}$ where the operator $A$ is applied.

\section{Proof System}\label{sec:proof}
In this section we present a proof system for local reasoning about probabilistic behaviour of quantum programs. We first define an assertion language, then propose a Hoare-style proof system, and finally prove the soundness of the system.

\subsection{Assertion Language}\label{sec:ass}                        
\begin{table}[t]
\caption{Syntax of assertion languages}
\label{tab:syn_ass}
\[
\begin{aligned}
	&(\mbox{Pure expression})&  \ &e& \  &::=& \  & n \ |\ x,y,...\ |\ f_{m}(e,...,e)\\
	&(\mbox{Pure formula})&   &P&  &::=& & \textbf{true} \ | \ \textbf{false} \ | \ P_{m}(e,...,e)\ |\ P\wedge P\ |\  \neg P \ | \  \forall x.P(x) \\
	&(\mbox{Quantum expression})&  &\ket{s}& &::=& &\ket{a}_{\overline{q}}\ |\ \ket{s}\otimes\ket{s} \\
  &(\mbox{State formula})& &F& &::=& & P\ |\ \ket{s}\ | \ F\odot F \ | \ F\wedge F
  % \ |\ \neg F%\ |\ \textcolor{teal}{  F} \\ %\textcolor{teal}{\ |\ \forall x.F(x)}
  \\
	&(\mbox{Distribution formula})& &D& &::=& &\oplus_{i\in I}p_{i}\cdot F_{i}\ |\ \oplus_{i\in I}F_{i}
\end{aligned}
\]
\end{table}

We now introduce an assertion language for our programs, whose syntax is given in Table~\ref{tab:syn_ass}.
Pure expressions are arithmetic expressions with integer constants, ranged over by $e$.
Pure formulas, ranged over by $P$, include the boolean constants $\mathbf{true}$ and $\mathbf{false}$, boolean predicates in the form $P_{m}$ and any $P$ connected by boolean operators such as negation, conjunction, and universal quantification. They are intended to  capture properties of classical states. Quantum expressions, ranged over by $\ket{s}$, include quantum states of the form $\ket{a}_{\overline{p}}$ and the tensor product $\ket{s}\otimes\ket{s}$ which we abbreviate as  $\ket{s}\ket{s}$. Here $\ket{a}_{\overline{p}}$ can be any computational basis or their linear combinations in the Hilbert space $\mathcal{H}_{\overline{p}}$. State formulas, ranged over by $F$, are used to express properties on both classical and quantum states, which include the pure formula $P$, the quantum expression $\ket{s}$ and any expression connected by boolean operator conjunction.
In addition, we introduce a new connective $\odot$ to express an assertion of two separable systems. Following \cite{Sun22}, we use free($F$) to denote the set of all free classical and quantum variables in $F$. For example, $\mbox{free}(|a_1\>_{\overline{q_1}}\otimes |a_2\>_{\overline{q_2}}) = \overline{q_1}\overline{q_2}$. Moreover, we use qfree($F$) to denote the set of all quantum variables in $F$. For the formula $F_{1} \odot F_{2}$ to be well defined, we impose the syntactical restriction that $\mathrm{qfree}(F_1)\cap\mathrm{qfree}(F_2)=\emptyset$.
Intuitively, a quantum state satisfies  $F_{1} \odot F_{2} $ if the state mentions two disjoint subsystems whose states satisfy  $F_{1}$ and $F_{2}$ respectively. 
Distribution formulas consist of some state formulas $F_{i}$ connected by the connective $\oplus$  with the weights given by $p_{i}$ satisfying $\sum_{i\in I}p_i=1$ as well as the non-probabilistic formula $\oplus_{i\in I} F_{i}$.
%It is worth noting that within a distribution formula each $F_{i}$ is required that $F_{i} = F_{j}$ holds if and only if $i= j$, where $i$, $j$ $\in$ $I$ and $\Sigma_{i \in I}p_{i} = 1$.
If there is a collection of distribution formulas $D_i=\oplus_j p_{ij}\cdot F_{ij}$ and a collection of probabilities $p_i$ with $\sum_{i\in I}p_i=1$, we sometimes write $\oplus_ip_i\cdot D_i$ to mean the formula $\oplus_{ij}p_i p_{ij}\cdot F_{ij}$.

\begin{table}[t]
\caption{Semantics of assertions}
\label{tab:sem_ass}
\[
\begin{array}{rllll}
	(\sigma,\rho) &\models & P \ &\mathrm{if}& \quad   [\![P]\!]_{\sigma}=\textbf{true}\\

  (\sigma,\rho) &\models& \ket{s}\ &\mathrm{if}& \quad  
	\frac{\rho}{\tr(\rho)}|_{\overline{q}}= |s\>\<s| \quad\mbox{where } \overline{q}=\mbox{free}(\ket{s}) \ \subseteq \mathrm{dom} (\rho) \\

        (\sigma,\rho) &\models& F_{1}\wedge F_{2}  \
	&\mathrm{if}& \quad  (\sigma,\rho) \models F_{1} 
	\wedge (\sigma,\rho) \models F_{2} \\

        (\sigma,\rho) &\models&   F_{1} \odot F_{2} \
	&\mathrm{if}& \quad   (\sigma,\rho|_{\mbox{qfree}(F_1)}) \models F_{1} \wedge (\sigma,\rho|_{\mbox{qfree}(F_2)}) \models F_{2}  \\ 
 & & & &   \quad \wedge \ \mbox{qfree}(F_1) \cap \mbox{qfree}(F_2) = \emptyset    \\

\mu & \models & F & \mathrm{if} & \quad \forall \, \sigma\in\support{\mu}. \ (\sigma,\mu(\sigma))\models F \\
        
	\mu &\models& \oplus_{i\in I}p_{i}\cdot F_{i} \  &\mathrm{if}& \quad \exists\,\mu_{1} \cdots \exists\,\mu_{m}.\  \\
 & & & & \quad [ \mu=\sum_{i\in I}p_{i}\cdot \mu_{i} \wedge (\forall i, p_i \neq 0 \rightarrow \size{\mu_i} = \size{\mu} \wedge \mu_{i} \models F_{i}) ] \\
    &&&& \quad \mbox{for } I=\{1,\ldots,m\} \\
	\mu &\models& \oplus_{i\in I} F_{i} \  &\mathrm{if}& \quad \exists\,p_{1} \cdots \exists p_{m}.\
    [(\bigwedge_{i \in I} p_{i} \ge 0) \wedge \sum_i p_i = 1  \wedge \mu \models \oplus_{i\in I}p_i\cdot F_i]\\ &&&& \quad \mbox{for } I=\{1,\ldots,m\}
\end{array}
\]
\end{table}

We use the notation $\mu \models F$ to indicate that the state $\mu$  satisfies the assertion $F$. %, which is interpreted 
The satisfaction relation $\models$ is defined in Table~\ref{tab:sem_ass}. When writing $(\sigma,\rho)\models F$, we mean that $(\sigma,\rho)$ is a machine state and $\rho$ is its quantum part representing the status of the whole quantum system in a program under consideration.
We use $[\![P]\!]_{\sigma}$ to denote the evaluation of the pure predicate $P$ with respect to  the classical state $\sigma$.

A machine state $(\sigma,\rho)$ satisfies the formula $|s\>$ if the reduced density operator obtained by first normalising $\rho$ and then restricting it to $\mbox{free}(|s\>)$ becomes $|s\>\<s|$, given that $\mathrm{qfree}(\ket{s})\subseteq \mathrm{dom}(\rho)$. The state $(\sigma,\rho)$ satisfies the formula $F_1\odot F_2$ if qfree$(F_1)$ and qfree$(F_2)$ are disjoint and the restrictions of $\rho$ to qfree($F_1$) and qfree($F_2$) satisfy the two sub-formulas $F_1$ and $F_2$, respectively. The assertion $F$ holds on a distribution $\mu$ when $F$ holds on each  state in the support of $\mu$.
A distribution state $\mu$ satisfies the distribution formula $\oplus_{i\in I}p_i\cdot F_i$ if $\mu$ is a linear combination of some $\mu_i$ with weights $p_i$ such that each $\mu_i$ has the same size as $\mu$ and satisfies $F_i$ whenever $p_i > 0$. 
 The formula $\oplus_{i\in I}F_i$ is a nondeterministic version of the distribution formulas in the form $\oplus_{i\in I}p_i\cdot F_i$ without fixing the weights $p_i$, so the weights can be arbitrarily chosen as long as their sum is $1$. For other assertions, the semantics should be self-explanatory.

Now we present several basic lemmas that form the foundation for our analysis of assertions within the given framework. These lemmas establish the soundness of our proof system and facilitate the manipulation of assertions.

\begin{lemma}
\label{lem:sat_res}
Let \(\mu\) be a distribution and \(F\) an assertion. For any \(V \subseteq \mathbf{QVar}\) such that \(\mathrm{qfree}(F) \subseteq V \subseteq \mathrm{dom}(\mu) \), we have that \(\mu \models F \Leftrightarrow \mu|_{V} \models F\).
\end{lemma}
This lemma indicates that any assertion $F$ is restrictive, enabling assertion verification by restricting attention to quantum variables referred to by \( F \).
\begin{lemma}
\label{lem:sat_linear}
The satisfaction relation is linear. Specifically, for any collection of probabilities $p_i$ 
with $0 < \sum_i p_i \le 1$ and distributions \( \mu_i \) such that \( \mu_i \models F \) whenever \( p_i > 0 \), it holds that $\sum_{i \in I} p_i \cdot \mu_i \models F$.
\end{lemma}

% \whl{This property guarantees that assertions are preserved under probabilistic mixtures of distributions, a critical feature for reasoning about quantum programs with probabilistic branches.}

% \whl{The following lemma is pivotal for proving the soundness of the [QFrame] rule presented in~\ref{sec:proofsys}.}

\begin{lemma}
\label{lem:sat-pre}
    Let \(\mu\) be a distribution and \(F\) an assertion. If \(\mu \models F\) and \(\mathrm{free}(F) \cap \mathrm{mod}(c) = \emptyset\), then \([\![c]\!]_{\mu} \models F\).
\end{lemma}

All the lemmas presented above can be proved by induction on the structure of assertions and programs. The difference between the two connectives \(\wedge\) and \(\odot\) is reflected by the following lemma, which is directly derivable from their semantic definitions. 
\begin{lemma}
\label{lem:sat_wedge_odot}
\begin{enumerate} 
    \item[(i)] \(\mu \models F_1 \wedge F_2 \Leftrightarrow \mu \models F_1 \wedge \mu \models F_2\);
    \item[(ii)] \(\mu \models F_1 \odot F_2 \Leftrightarrow \mu \models F_1 \wedge \mu \models F_2 \wedge \mathrm{qfree}(F_1) \cap \mathrm{qfree}(F_2) = \emptyset\).
\end{enumerate}
\end{lemma}

The next lemma shows that the semantics for \(\odot\) aligns with the conventional understanding of separable conjunction.

\begin{lemma}
\label{lem:seman-odot-eq}
Let \(\sigma\) be a classical state, \(\rho\) a quantum state, and \(F_1, F_2\) two assertions. Then \((\sigma, \rho) \models F_1 \odot F_2\) if and only if there exist \(V_1, V_2, \rho_1, \rho_2\) such that \(V_1 \cap V_2 = \emptyset\), \(\rho_1 \in \padist{\CH_{V_1}}\), \(\rho_2 \in \padist{\CH_{V_2}}\), and
\[
\rho|_{V_1 \cup V_2} = \rho_1 \otimes \rho_2 \wedge (\sigma,\rho_1) \models F_1 \wedge (\sigma,\rho_2) \models F_2.
\]
\end{lemma}

Detailed proofs of these lemmas are formalized in Coq. Finally, the following proposition is important for proving the soundness of the [While] rule presented in Subsection~\ref{sec:proofsys}. Its proof proceeds by induction on the structure of \( D\), and the details are presented in Appendix~\ref{sec:closed}.

\begin{proposition}\label{prop:closed}
Let \(V \subseteq \mathbf{QVar}\). For any assertion \(D\), the set \([\![D]\!]_{V} \triangleq \{ \mu \ | \ \mu \models D \wedge \mathrm{dom}(\mu)=V \} \) is a closed set.
\end{proposition}

From the  relation $\models$, we can derive a quantitative definition of satisfaction, where a distribution satisfies a predicate only to certain degree.

\begin{definition}
Let $F$ be an assertion and $p\in [0,1]$ a real number. We say that the probability of a distribution $\mu$ satisfying $F$ is $p$, written by
\[\mathbb{P}_{\mu}(F)=p,\]
if there exist two distributions $\mu_{1}$and $\mu_{2}$ such that 
$\mu=p\mu_{1}+(1-p)\mu_{2}$, $\mu_{1} \models F$ and $\mu_{2} \models \neg F$, and moreover, $p$ is the maximum probability for this kind of decomposition of $\mu$.
\end{definition}

%\paragraph{Distribution formulas}
 In \cite{Sun22}, assertions with disjunctions are used as the postconditions of measurement statements. However, this approach is awkward when reasoning about probabilities. Let us take a close look at the problem in Example~\ref{ex3}.

\begin{example}
  \label{ex3}
  We revisit the program $\textbf{addM}$ discussed in Example~\ref{ex1}, which measures the variables $q_0$ and $q_1$, both in the state $|+\>\<+|$,
	and subsequently adds their results of measurements. Here $M$ is a projective  measurement $\{|0\>\<0|,  |1\>\<1|\}$.
	
	\[
	\begin{aligned}
		\mathbf{addM} \ \triangleq\quad  
		&q_{0}:=\ket{0}; H[q_0];\\
		&q_{1}:=\ket{0}; H[q_1];\\
		&v_{0}:=M[q_{0}];\\
		&v_{1}:=M[q_{1}];\\
		&v:=v_{0}+v_{1}\\
	\end{aligned}
	\]
        After the measurements on $q_0$ and $q_1$, the classical variables $v_0$ and $v_1$  are assigned to either $0$ or $1$ with equal probability. By executing the last command,  we assign the sum of $v_0$ and $v_1$ to $v$, and obtain the following POVD $\mu$.
         \[\begin{array}{lclll}
     &   v_0v_1v & & & q_0q_1\\
     &   000 & \mapsto & \tfrac{1}{4} & |00\>\<00|\\
 \mu: &   011 & \mapsto & \tfrac{1}{4} & |01\>\<01|\\
     &   101 & \mapsto & \tfrac{1}{4} & |10\>\<10|\\
     &   112 & \mapsto & \tfrac{1}{4} & |11\>\<11|
        \end{array}\]
         The second column represents the four classical states while the last column shows the four quantum states.
         For example, we have $\sigma_1(v_0v_1v)=000$ and $\rho_1=|00\>\<00|$. Let $\mu_i=(\sigma_i,\rho_i)$ for $1\leq i\leq 4$.
         Then $\mu=\tfrac{1}{2}\mu_{14}+\tfrac{1}{2}\mu_{23}$,
         where $\mu_{14}=\tfrac{1}{2}\mu_1+\tfrac{1}{2}\mu_4$ and $\mu_{23}$=$\tfrac{1}{2}\mu_2+\tfrac{1}{2}\mu_3$.
         Since $\mu_i\models (v=1)$ for $i=2,3$, it follows that
         $\mu_{23}\models (v=1)$ by Lemma~\ref{lem:sat_linear}. On the other hand, we have $\mu_{14}\models (v\neq 1)$.
         Therefore, it follows that $\mathbb{P}_\mu(v=1)=\tfrac{1}{2}$.
         That is, the probability for $\mu$ to satisfy the assertion $v=1$ is $\tfrac{1}{2}$.
         
         Alternatively, we can express the above property as a distribution formula. Let $D=\frac{1}{2}\cdot (v=1)\oplus\frac{1}{2}\cdot (v\neq 1)$. We see that $\mu\models D$ due to the fact that $\mu=\frac{1}{2}\mu_{23}+\frac{1}{2}\mu_{14}$, $\mu_{23}\models (v=1)$ and $\mu_{14}\models (v\neq 1)$.
         
In~\cite{Sun22}, there is no distribution formula.
The best we can do is to use a disjunctive assertion to describe the postcondition of the above program.
	\[
	\begin{aligned}
		F \triangleq\ & \tfrac{1}{4} \cdot (v=0\wedge \ket{00}_{q_{0}q_1}) \vee \tfrac{1}{4}\cdot (v=1\wedge \ket{01}_{q_{0}q_1}) \ \vee\\
        & \tfrac{1}{4} \cdot(v=1\wedge \ket{10}_{q_{0}q_1}) \vee \tfrac{1}{4} \cdot(v=2\wedge \ket{11}_{q_{0}q_1}).
	\end{aligned}
	\]
This assertion does not take the mutually exclusive correlations between different branches into account.
For example, it is too weak for us to prove that $\mathbb{P}(v=1)=\frac{1}{4}+\frac{1}{4}=\frac{1}{2}$, %with more details referring to 
as discussed in more details in~\cite{Sun22}.
From this example, we see that distribution formulas give us a more accurate way of describing the behaviour of measurement statements than disjunctive assertions.
\end{example}

\subsection{Proof System}\label{sec:proofsys}
In this subsection, we present a series of inference rules for classical--quantum programs, which will be proved to be sound at the end of this subsection. As usual, we use the Hoare triple $\{D_1\}$ $c$ $\{D_2\}$ to express the correctness of our programs, where $c$ is a program and $D_1$, $D_2$ are the assertions specified in Table~\ref{tab:syn_ass}.

\begin{table}[htb]
\caption{Inference rules for classical statements}
\label{tab:rule_I}
\centering
\[\begin{array}{c}
	\dfrac{}{\{D\} \ \textbf{skip} \ \{ D\}} [\mathrm{Skip}]  \qquad 
	\dfrac{}{\{D\} \ \textbf{abort} \ \{ \textbf{false}\} } [\mathrm{Abort}]\\
	\dfrac{}{\{D[a/x]\}\ x:=a\ \{ D\}} [\mathrm{Assgn}]
	\qquad \dfrac{\{D_{0}\}\ c_{0} \ \{ D_{1}\} \quad \{D_{1}\} \ c_{1}\ \{ D_{2}\} }{\{D_{0}\} \ c_{0};c_{1} \{D_{2}\}} [\mathrm{Seq}]\\

	\dfrac{\{ F_{1} \wedge   b  \} \ c_{1}\ \{ F'_{1}\}  \quad \{ F_{2}\wedge \neg b  \} \ c_{2} \ \{ F'_{2}\} }{\{ p(F_{1}\wedge b )\oplus (1-p) (F_{2} \wedge   \neg b )\} \ \mathbf{if} \ b \ \mathbf{then} \ c_{1}\  \mathbf{else} \ c_{2}\ {\bf fi} \ \{pF'_{1}\oplus (1-p) F'_{2} \}}	[\mathrm{Cond}]	\\
	\dfrac{}{\{\textbf{false}\} \ c \ \{D\}}[\mathrm{Absurd}] 
	\qquad 
	\dfrac{D_{0}\Rightarrow D_{1} \quad \{D_{1}\} \ c \ \{D_{2}\} \quad D_{2}\Rightarrow D_{3}}{\{D_{0}\}  \ c \ \{D_{3}\}} [\mathrm{Conseq}]\\
 
	\dfrac{ D=(F_{0}\wedge   b  )\oplus (F_{1} \wedge   \neg b ) \quad \{ F_{0} \wedge   b  \}  \ c \ \{ D\}}{\{D \} \ \mathbf{while} \ b \  \mathbf{do} \ c \ {\bf od} \ \{ F_{1} \wedge     \neg b \}}[\mathrm{While}]\\
	
	\dfrac{\{F_1\} \ c \ \{F_1'\} \quad \{F_2\} \ c\  \{F'_2\}}{\{F_1\wedge F_2\} \  c \ \{F'_1\wedge F'_2\}}[\mathrm{Conj}] \\
        	\dfrac{\{F_1\} \ c\ \{F_2\} \quad  \mathrm{free}(F_3) \cap \mathrm{mod}(c)=\emptyset}{\{F_1 \odot F_3\}\  c \ \{F_2 \odot F_3\}}[\mathrm{QFrame}]\\
    
	\dfrac{\forall i\in I. \ \{D_{i}\} \ c \ \{ D'_{i}\} \quad \sum_{i\in I}p_{i}= 1}{ \{\oplus_{i\in I} p_{i}\cdot D_{i}\} \ c \ \{\oplus_{i\in I} p_{i}\cdot D'_{i}\} }[\mathrm{Sum}]
\end{array}\]
\end{table}

Table~\ref{tab:rule_I} %captures 
lists the rules for classical statements, with most of them being standard and thus self-explanatory. The assertion $\mathit{D}[a/x]$ 
is the same as $D$ except that all the free occurrences of variable $x$ in $D$ are replaced by $a$. The assertion $\mathit{D}_1$$\Rightarrow$$\mathit{D}_2$  indicates that $\mathit{D_1}$ logically implies $\mathit{D}_2$. 
When all the formulas involved in the [While] and [Cond] rules presented in Table~\ref{tab:rule_I} are pure formulas, they can deduce their corresponding classical counterparts, respectively.
The quantum frame rule [QFrame] is introduced for local reasoning, which allows us to add assertion $F_3$ to the pre/post-conditions of the local proof $\{F_1\}$ $c$ $\{F_2\}$.  The condition $\free(F_3) \cap \modi(c)=\emptyset$ indicates that all the free classical and quantum variables in $F_3$ are not modified by program $c$. Note that when $F_3$ is a pure assertion, $\odot $ can be replaced by $\wedge$.
 The rule [Sum] allows us to reason about a probability distribution by considering each pure state individually.

\begin{table}[t]
\caption{Inference rules for quantum statements}
\label{tab:rule_Q}
\begin{gather*}
	\dfrac{}{\{\textbf{true}\}\  \overline{q} :=\ket{0} \ \{ \ket{0}_{\overline{q}}\}}[\mathrm{QInit}]
	\qquad 
	\dfrac{}{\{U_{\overline{q}}^{\dagger}F\} \ U[\overline{q}] \ \{ F\}}[\mathrm{QUnit}]\\
	\dfrac{M=\{M_{i}\}_{i\in I} \quad p_{i}=\Vert {M_{i}}_{\overline{q}}\ket{v}\Vert^{2}}{\{\wedge_i (P_{i}[i/x] \wedge \ket{v}_{\overline{q}\overline{q'}})\} \ x:=M[\overline{q}] \
    \{\oplus_{i} p_{i}\cdot (  P_{i} \wedge ({M_{i}}_{\overline{q}}\ket{v}_{\overline{q}\overline{q'}}/\sqrt{p_{i}}))\}}[\mathrm{QMeas}]		
\end{gather*}

\end{table}
Table~\ref{tab:rule_Q} displays the inference rules for quantum statements. In rule [QInit], we see that the execution of the command $\overline{q} :=\ket{0}_{\overline{q}}$ sets the quantum system $\overline{q}$ to $|0\>$,  no matter what the initial state is. In rule [QUnit], for the postcondition $F$ we have the precondition $U^{\dagger}_{\overline{q}}F$. Here, $U^{\dagger}_{\overline{q}}$ distributes over those connectives of state formulas and eventually applies to quantum expressions. For example, if $F=\ket{v}_{\overline{q}\overline{q'}}\wedge P$, then $U^{\dagger}_{\overline{q}}F=U^{\dagger}_{\overline{q}}(\ket{v}_{\overline{q}\overline{q'}})\wedge P$. 
In rule [QMeas], the combined state of the variables $\overline{q}\overline{q'}$ is specified because there may be an entanglement between the subsystems for $\overline{q}$ and $\overline{q'}$. In that rule, we write $P_{i}[i/x]$ for the assertion obtained from $P_{i}$ by replacing the variable $x$ with value $i$.  The postcondition is a distribution formula, with  each assertion $P_{i} \wedge ({M_{i}}_{\overline{q}}\ket{v}_{\overline{q}\overline{q'}}/\sqrt{p_{i}})$ assigned probability $p_i$,
i.e., the probability of obtaining outcome $i$ after the measurement $M$.

\begin{table}[t]
\caption{Inference rules for entailment reasoning}
\label{tab:rule_E}
  \begin{gather*}
 \dfrac{}{F \vdash \true}[\mathrm{PT}] \qquad
 \dfrac{}{F \odot \true \dashv\vdash F}[\mathrm{OdotE}]\\ 
 \dfrac{}{F_1 \odot F_2 \dashv\vdash F_2 \odot F_1}[\mathrm{OdotC}] \qquad 
	\dfrac{}{F_1\odot (F_2 \odot F_3) \dashv\vdash (F_1 \odot F_2) \odot F_3}[\mathrm{OdotA}]\\ 
% \dfrac{F_1 \vdash F_2}{F_1\odot F_3 \vdash F_2 \odot F_3}[\mathrm{OdotV}] \qquad 
 \dfrac{}{P_1 \odot P_2 \dashv\vdash P_1 \wedge P_2 }[\mathrm{OdotO}] \qquad
 \dfrac{}{P \odot F \dashv\vdash P \wedge F }[\mathrm{OdotOP}] \\ 
 %\quad  \dfrac{}{p \odot Q \dashv\vdash pQ }[\mathrm{OdotP}] \\
 \dfrac{}{P\wedge (F_1 \odot F_2) \dashv\vdash (P \wedge F_1) \odot F_2}[\mathrm{OdotOA}]\\ 
 \dfrac{}{F_1 \odot (F_2 \wedge F_3) \dashv\vdash (F_1\odot F_2) \wedge (F_1\odot F_3)}[\mathrm{OdotOC}] \\ 
% \dfrac{}{ F_1\odot(F_2\oplus ) \dashv\vdash (P\odot Q) \oplus ( P \odot R)} [\mathrm{OdotOF}]\\ \\
 \dfrac{}{\ket{u}_{\overline{p}} \ket{v}_{\overline{q}}\dashv\vdash  \ket{v}_{\overline{q}} \ket{u}_{\overline{p}}}[\mathrm{ReArr}]\quad
 \dfrac{}{\ket{u}_{\overline{p}} \ket{v}_{\overline{q}} \dashv\vdash \ket{u\otimes v}_{\overline{p}\overline{q}}}[\mathrm{Separ}]\\ 
 \dfrac{}{\ket{u}_{\overline{p}} \ket{v}_{\overline{q}} \dashv\vdash \ket{u}_{\overline{p}} \odot \ket{v}_{\overline{q}}}[\mathrm{OdotT}] \\
 \dfrac{}{p_0\cdot F \oplus p_1\cdot F \oplus p_2 \cdot F' \dashv\vdash (p_0+p_1)\cdot F \oplus p_2\cdot F'}[\mathrm{OMerg}]\\
 \dfrac{}{\oplus_{i\in I}p_i\cdot F_i \vdash \oplus_{i\in I} F_i}[\mathrm{Oplus}] \qquad
 \dfrac{\forall i\in I,\ F_i\vdash F'_i}{\oplus_{i\in I}p_i\cdot F_i \vdash \oplus_{i\in I}p_i\cdot F'_i}[\mathrm{OCon}]
\end{gather*}
\end{table}

Table~\ref{tab:rule_E} presents several rules for entailment reasoning about quantum predicates. The notation $D_1\vdash D_2$ says that $D_1$ proves $D_2$. Intuitively, it means that any state satisfying $D_1$ also satisfies $D_2$. We write $D_1\dashv\vdash D_2$ if the other direction also holds.

The connective $\odot$ is commutative and associative, according to the rules [OdotC] and [OdotA].  
If one or two assertions are pure, the rules [OdotO] and [OdotOP] replace $\odot$ with $\wedge$.
The rule [OdotOA] replaces $P \wedge (F_1 \odot F_2) $ with $(P \wedge F_1) \odot F_2$ and vice versa. The rule [OdotOC]  assists us to distribute $\odot$ into conjunctive  assertions.
The rule [ReArr] allows us to rearrange quantum expressions while [Separ] allows us to split/join the quantum expressions, given that $\overline{p}$ and $\overline{q}$ are not entangled with each other.
The rules [ReArr] and [Separ] can be obtained naturally via the properties of tensor products. The rule [OdotT] replaces the $\odot$ connective with $\otimes$ when both assertions are state expressions.
The rule [OMerg] allows us to merge the probabilities of two branches in a distribution formula if the two branches are the same.
The rule [Oplus] is easy to be understood as $\oplus_{i\in I} F_i$ is essentially a relaxed form of $\oplus_{i\in I}p_i\cdot F_i$. The rule [OCon] says that if each $F_i$ entails $F'_i$, then the entailment relation is preserved between the combinations $\oplus_{i\in I}p_i\cdot F_i$ and $\oplus_{i\in I}p_i\cdot F'_i$.
We use the notation $\vdash \{D_1\} \ c \ \{D_2\}$ to mean that
the Hoare triple $\{D_1\} \ c \ \{D_2\}$ is provable by applying the rules in Tables~\ref{tab:rule_I}--\ref{tab:rule_E}.

\begin{example}
\label{exam2}
	Suppose there are two separable systems $\bar{q}$ and $\bar{q'}$. They satisfy the precondition  $\frac{1}{2}\ket{u_{1}}_{\bar{q}}\ket{v_{1}}_{\bar{q'}}\oplus \frac{1}{2}\ket{u_{2}}_{\bar{q}}\ket{v_{2}}_{\bar{q'}}$. After applying the operator $\mathit{U}[\bar{q}]$, they satisfy the postcondition  $\frac{1}{2}(U_{\bar{q}}\ket{u_{1}}_{\bar{q}})\ket{v_{1}}_{\bar{q'}}\oplus \frac{1}{2}(U_{\bar{q}}\ket{u_{2}}_{\bar{q}})\ket{v_{2}}_{\bar{q'}}$. In other words, the following Hoare triple holds. 
\begin{equation}\label{eq0}
	\begin{aligned}
   \vdash\ & \Key{\{\tfrac{1}{2}\ket{u_{1}}_{\bar{q}}\ket{v_{1}}_{\bar{q'}}\oplus \tfrac{1}{2}\ket{u_{2}}_{\bar{q}}\ket{v_{2}}_{\bar{q'}}\}}\\
	&U[\bar{q}]\\
	&\Key{\{\tfrac{1}{2}(U_{\bar{q}}\ket{u_{1}}_{\bar{q}})\ket{v_{1}}_{\bar{q'}} \oplus \tfrac{1}{2}(U_{\bar{q}}\ket{u_{2}}_{\bar{q}})\ket{v_{2}}_{\bar{q'}}\}}
   \end{aligned}
\end{equation}
This Hoare triple can be proved as follows. Firstly, we apply the rule [QUnit] to obtain
\begin{equation}\label{eq1}
          \vdash\ \Key{\{|u_1\>_{\overline{q}}\}}\ U[\overline{q}] \ \Key{\{U_{\overline{q}}|u_1\>_{\overline{q}}\}}.
          \end{equation}
Then we use the rules [QFrame] and [OdotT] to get
        \begin{equation}\label{eqx}
          \vdash\ \Key{\{|u_1\>_{\overline{q}}|v_1\>_{\overline{q'}}\}}\ U[\overline{q}] \ \Key{\{(U_{\overline{q}}|u_1\>_{\overline{q}})|v_1\>_{\overline{q'}}\}}.
         \end{equation}
        Similarly, we have
        \begin{equation}\label{eqy}
        \vdash\  \Key{\{|u_2\>_{\overline{q}}|v_2\>_{\overline{q'}}\}}\ U[\overline{q}] \ \Key{\{(U_{\overline{q}}|u_2\>_{\overline{q}})|v_2\>_{\overline{q'}}\}}.
        \end{equation}
        Combining (\ref{eqx}) with (\ref{eqy}) by rule [Sum], we obtain the Hoare triple in (\ref{eq0}).
       
\end{example}

\begin{example}
  Let us consider the program \textbf{addM} and the distribution formula $D=\frac{1}{2}\cdot (v=1)\oplus\frac{1}{2}\cdot (v\neq 1)$ discussed in Example~\ref{ex3}. Now we would like to formally prove that
  \[\Key{\{\true\}}\ \textbf{addM}\ \Key{\{D\}} \ .\]
The proof outline is given in Table~\ref{t:addm}. Following \cite{Sun22}, we use the following notations to highlight the application of the frame rule [QFrame]:
 \[\InOut\quad \Key{\{\ F_1\ \}}\ c \ \Key{\{\ F_2\ \}} \qquad \mbox{or equivalently} \qquad \begin{array}{lc}
   \In & \Key{\{\ F_1\ \}} \\
   & c \\
   \Out & \Key{\{\ F_2\ \}} 
   \end{array}\]
 Both notations indicate that $\{F_1\} c \{F_2\}$ is a local proof for $c$ and is useful for long proofs.
 The frame assertion $F_3$ can be deduced from the assertions before $F_1$ or after $F_2$. 
\end{example}

\begin{table}[t]
 \caption{Proof outline for the \textbf{addM} program}
 \label{t:addm}
 \begin{tabular}{l}
  $\Key{\{\ \true\ \}}$\\
  $\Key{\{\ \true \odot \true \ \} \qquad\mbox{by rule [OdotE]}}$\\
  $\InOut \Key{\{\  \mathbf{true}\ \}} \quad  q_0:=\ket{0}; \quad \Key{\{\  \ket{0}_{q_0} \ \} \qquad\mbox{by rule [QInit]}}$\\
  $\InOut \Key{\{\ \ket{0}_{q_0}\ \}} \quad  H[q_0]; \quad \Key{\{\  \ket{+}_{q_0} \ \} \qquad\mbox{by rule [QUnit]}}$\\
   $\InOut \Key{\{\  \mathbf{true}\ \}} \quad  q_1:=\ket{0}; \quad \Key{\{\  \ket{0}_{q_1} \ \} \qquad\mbox{by rule [QInit]}}$\\
  $\InOut \Key{\{\ \ket{0}_{q_1}\ \}} \quad  H[q_1]; \quad \Key{\{\  \ket{+}_{q_1} \ \} \qquad\mbox{by rule [QUnit]}}$\\
   $\Key{\{\ |++\>_{q_0q_1} \wedge (v_0=0)[0/v_0]\wedge (v_0=1)[1/v_0]\ \}    \qquad\mbox{by rule [QFrame] and [Conseq]}} $ \\
   $v_0:=M[q_0];$\\
   $\Key{\{\ \frac{1}{2}\cdot (|0+\>_{q_0q_1}\wedge v_0=0) \oplus \frac{1}{2}\cdot (|1+\>_{q_0q_1}\wedge v_0=1) \ \}\qquad\mbox{by rule [QMeas]}}$ \\
   $\Key{\{\ \frac{1}{2}\cdot (|0+\>_{q_0q_1}\wedge v_0=0 \wedge (v_1=0)[0/v_1]\wedge (v_1=1)[1/v_1])}$\\
   $\Key{\oplus \frac{1}{2}\cdot (|1+\>_{q_0q_1}\wedge v_0=1\wedge (v_1=0)[0/v_1]\wedge (v_1=1)[1/v_1]) \ \}\qquad\mbox{by rule [Conseq]}}$ \\
   $v_1:=M[q_1];$\\
   $\Key{\{\ \frac{1}{4}\cdot (|00\>_{q_0q_1}\wedge v_0=0 \wedge v_1=0)}$\\
   $\Key{\oplus \frac{1}{4}\cdot (|01\>_{q_0q_1}\wedge v_0=0 \wedge v_1=1)}$\\
   $\Key{\oplus \frac{1}{4}\cdot (|10\>_{q_0q_1}\wedge v_0=1\wedge v_1=0)}$\\
   $\Key{\oplus \frac{1}{4}\cdot (|11\>_{q_0q_1}\wedge v_0=1\wedge v_1=1) \ \}\qquad\mbox{by rules [QMeas] and [Sum]}}$ \\
     $\Key{\{\ \frac{1}{4}\cdot (|00\>_{q_0q_1}\wedge v_0=0 \wedge v_1=0 \wedge (v=0)[(v_0+v_1)/v])}$\\
   $\Key{\oplus \frac{1}{4}\cdot (|01\>_{q_0q_1}\wedge v_0=0 \wedge v_1=1\wedge (v=1)[(v_0+v_1)/v])}$\\
   $\Key{\oplus \frac{1}{4}\cdot (|10\>_{q_0q_1}\wedge v_0=1\wedge v_1=0\wedge (v=1)[(v_0+v_1)/v])}$\\
   $\Key{\oplus \frac{1}{4}\cdot (|11\>_{q_0q_1}\wedge v_0=1\wedge v_1=1\wedge (v=2)[(v_0+v_1)/v]) \ \}\qquad\mbox{by rule [Conseq]}}$ \\
   $v :=v_0+v_1$ \\
     $\Key{\{\ \frac{1}{4}\cdot (|00\>_{q_0q_1}\wedge v_0=0 \wedge v_1=0 \wedge v=0)}$\\
   $\Key{\oplus \frac{1}{4}\cdot (|01\>_{q_0q_1}\wedge v_0=0 \wedge v_1=1\wedge v=1)}$\\
   $\Key{\oplus \frac{1}{4}\cdot (|10\>_{q_0q_1}\wedge v_0=1\wedge v_1=0\wedge v=1)}$\\
   $\Key{\oplus \frac{1}{4}\cdot (|11\>_{q_0q_1}\wedge v_0=1\wedge v_1=1\wedge v=2) \ \}\qquad\mbox{by rule [Assgn] and [Sum]}}$ \\
   $\Key{\{\ \frac{1}{4}\cdot (v=0) \oplus \frac{1}{4}\cdot (v=1) \oplus \frac{1}{4}\cdot (v=1) \oplus \frac{1}{4}\cdot (v=2) \ \} \qquad\mbox{by rule [OCon]}}$\\
   $\Key{\{\ \frac{1}{4}\cdot (v\ne 1) \oplus \frac{1}{4}\cdot (v=1) \oplus \frac{1}{4}\cdot (v=1) \oplus \frac{1}{4}\cdot (v\ne 1) \ \} \qquad\mbox{by rule [OCon]}}$\\
$\Key{\{\ \frac{1}{2}\cdot (v=1)\oplus\frac{1}{2}\cdot (v\neq 1)\ \}\qquad\mbox{by rule [OMerg]}}$ \\
 \end{tabular}
\end{table}
%\subsection{Soundness}

Our inference system is sound in the following sense: any Hoare  triple in the form $\{D_1\} \ c \ \{D_2\}$ derived from the inference system is valid,  denoted by $\vDash \{D_1\} c \{D_2\}$, meaning that for any state $\mu$ we have  $\mu\models D_1$ implies $\denote{c}_\mu\models D_2$.

\setcounter{theorem}{0}
\begin{theorem}[Soundness]
For any trace-preserving program $c$,
    if $\vdash \{D_1\} \ c\ \{ D_2\}$ then $\vDash \{D_1\} \ c \ \{ D_2\}$.
    \label{sound}
\end{theorem}
% A detailed proof of Theorem~\ref{sound} is given in Appendix~\ref{appA}
\begin{proof}
The proof is carried out by rule induction. We focus on some difficult rules, [Cond], [While] and [QFrame], while omitting the others. Readers interested in the full proof details are invited to consult our Coq formalization.

Consider the rule [Cond] first.
\[
\dfrac{\{ F_{0} \wedge   b  \}\ c_{0}\ \{ F'_{0}\}  \qquad \{ F_{1}\wedge   \neg b  \} \ c_{1} \ \{ F'_{1}\} }{\{ p(F_{0}\wedge   b )\oplus (1-p) (F_{1} \wedge   \neg b )\} \ \mathbf{if} \ b \ \mathbf{then} \ c_{0}\   \mathbf{else} \  c_{1}\ \mathbf{fi} \ \{pF'_{0}\oplus (1-p) F'_{1} \}}
\]
Assume that $\mu \models p(F_{0}\wedge   b )\oplus (1-p) (F_{1} \wedge   \neg b )$.
Then there exist some states $\mu_{0}$ and  $\mu_{1}$ such that $\mu=p\mu_{0}+(1-p)\mu_{1}$, $\Vert \mu_0 \Vert = \Vert \mu_1 \Vert= \Vert \mu \Vert$, $\mu_{0}\models F_{0} \wedge    b $ and $\mu_{1}\models F_{1} \wedge   \neg b $.
By induction, the premises $\{ F_{0} \wedge   b  \}\ c_{0}\ \{ F'_{0}\}$ and $\{ F_{1}\wedge   \neg b\} \ c_{1}\ \{ F'_{1}\}$ are valid. It follows that
$[\![c_{0}]\!]_{\mu_{0}} \models F'_{0}$ and 
$[\![c_{1}]\!]_{\mu_{1}} \models F'_{1}$. Since the semantic function $\denote{\cdot}$ is linear,  we then have 
\[
\begin{aligned}
	& [\![\mathbf{if} \ b \ \mathbf{then} \ c_{0} \ \mathbf{else} \ \ c_{1} \ \mathbf{fi}]\!]_{\mu}\\
	=\ & p[\![\mathbf{if} \ b \ \mathbf{then} \ c_{0} \ \mathbf{else} \  \ c_{1} \ \mathbf{fi}]\!]_{\mu_{0}}
	+(1-p)[\![\mathbf{if} \ b \ \mathbf{then} \ c_{0} \ \mathbf{else} \  \ c_{1} \ \mathbf{fi}]\!]_{\mu_{1}} \\
	=\ & p[\![c_{0}]\!]_{\mu_{0}}+(1-p)[\![c_{1}]\!]_{\mu_{1}}.
\end{aligned}
\]
Since we consider programs that are trace-preserving,
it is easy to see that $p[\![c_{0}]\!]_{\mu_{0}}+(1-p)[\![c_{1}]\!]_{\mu_{1}} \models pF'_{0}\oplus (1-p)F'_{1}$. 

Now let us consider the rule [While].

\[
\dfrac{ D=(F_{0}\wedge  b)\oplus (F_{1} \wedge   \neg b) ,\quad \{ F_{0} \wedge   b  \}  \ c \ \{ D\}}{\{D \} \ \mathbf{while} \ b \ \mathbf{do} \ c \ \mathbf{od} \ \{ F_{1} \wedge     \neg b  \}}
\]
Suppose $\mu \models D$ for some state $\mu$. Then there exist states $\mu_0$, $\mu_1$ and a probability $p$  such that $\mu=p\mu_{0}+(1-p)\mu_{1}$, $\Vert \mu_0 \Vert = \Vert \mu_1 \Vert= \Vert \mu \Vert$, $\mu_0\models F_0\wedge   b$ and  $\mu_{1}\models F_{1} \wedge   \neg b $. By induction, the premise $ \{ F_{0} \wedge   b  \}  \ c\  \{ D\}$ is valid. It follows that $[\![c]\!]_{\mu_{0}} \models D$.

For any $n\geq 0$, we write $\mathbf{while}^n$ for the $n$-th iteration of the while loop, i.e. \[(\mathbf{if} \ b \ \mathbf{then} \ c \ \mathbf{fi})^{n};\mathbf{if} \ b \ \mathbf{then} \ \mathbf{abort} \ \mathbf{fi}.\]
We claim that if $\mu\models D$ then $[\![\mathbf{while}^n]\!]_\mu \models F_1\wedge \neg b$ for any $n\geq 0$.
This can be proved by induction on $n$.

\begin{itemize}
\item $n=0$. In this case we have that
\[\begin{array}{rcl}
\denote{\mathbf{while}^0}_\mu & = &
\denote{\mathbf{if} \ b \ \mathbf{then} \ \mathbf{abort} \ \mathbf{fi}}_\mu \\
& = & p \denote{\mathbf{if} \ b \ \mathbf{then} \ \mathbf{abort} \ \mathbf{fi}}_{\mu_0} + (1-p)\denote{\mathbf{if} \ b \ \mathbf{then} \ \mathbf{abort} \ \mathbf{fi}}_{\mu_1} \\
& = & p\varepsilon + (1-p)\mu_1\\
& = & (1-p)\mu_1.
\end{array}\]
Since $\mu_{1}\models F_{1} \wedge   \neg b $, it implies that $(1-p)\mu_{1}\models F_{1} \wedge \neg b$ by Lemma~\ref{lem:sat_linear}.

\item $n=k+1$. We infer that
\[\begin{array}{rcl}
\denote{\mathbf{while}^{k+1}}_\mu & = &
\denote{\mathbf{if} \ b \ \mathbf{then} \ c \ \mathbf{fi};\ \mathbf{while}^k}_\mu \\
& = & p \denote{\mathbf{if} \, b \, \mathbf{then} \, c \, \mathbf{fi};\ \mathbf{while}^k}_{\mu_0}
 + (1-p) \denote{\mathbf{if} \, b \, \mathbf{then} \, c \, \mathbf{fi};\ \mathbf{while}^k}_{\mu_1} \\
& = & p \denote{\mathbf{while}^k}_{\denote{c}_{\mu_0}} 
 + (1-p) \denote{\mathbf{while}^k}_{\mu_1}\\
& = &  p \denote{\mathbf{while}^k}_{\denote{c}_{\mu_0}} 
 + (1-p) \mu_1 .
\end{array}\]
Note that $\denote{c}_{\mu_0}\models D$.
Therefore, it follows from induction hypothesis that  $\denote{\mathbf{while}^k}_{\denote{c}_{\mu_0}} \models F_1\wedge \neg b$.
% As in the last case, we have $(1-p)\mu_1\models F_1\wedge \neg b$.
Since $\mu_{1}\models F_{1} \wedge   \neg b $, we immediately have that $\denote{\mathbf{while}^{k+1}}_\mu\models F_1\wedge \neg b $ by Lemma~\ref{lem:sat_linear}.
\end{itemize} 

Thus we have completed the proof of the claim. 
Note that if the domain of \(\mu\) is \(V \subseteq \mathbf{QVar}\), then the domain of \(\denote{\mathbf{while}^n}_\mu\) remains \(V\) for all \(n \ge 0\). Consequently, by Proposition~\ref{prop:closed}, we have 
\[
\denote{\mathbf{while} \ b \ \mathbf{do} \ c \ \mathbf{od}}_{\mu} = \lim_{n \rightarrow \infty} \denote{\mathbf{while}^n}_\mu \models F_1 \wedge \neg b. 
\]

Finally, we consider the rule [QFrame]: \\
 \[ \dfrac{\{F_1\} \ c\ \{F_2\} \quad  \mathrm{free}(F_3) \cap \mathrm{mod}(c)=\emptyset}{\{F_1 \odot F_3\}\  c \ \{F_2 \odot F_3\}}[\mathrm{QFrame}].\]
  Consider a distribution \(\mu\) such that \(\mu \models F_1 \odot F_3\). By Lemma~\ref{lem:sat_wedge_odot}, we can infer that \(\mu \models F_1\) and \(\mu \models F_3\). Define \(\mu' = \denote{c}_{\mu}\). Suppose that \(\mathrm{qfree}(F_2) \cap \mathrm{qfree}(F_3) = \emptyset\). Our objective is to demonstrate that:
\[
\mu' \models F_2 \quad \text{and} \quad \mu' \models F_3.
\]
By induction, the premise \(\{F_1\} \ c\ \{F_2\}\) is valid. We immediately obtain \(\mu' \models F_2\). Since \(\mathrm{free}(F_3) \cap \mathrm{mod}(c) = \emptyset\), it follows from Lemma~\ref{lem:sat-pre} that \(\mu' \models F_3\). Consequently, by invoking Lemma~\ref{lem:sat_wedge_odot}, we conclude that \(\mu' \models F_2 \odot F_3\). \qedhere

\end{proof}

\section{Case Analysis}\label{sec:cases}
Here, we apply the proof system to verify the functional correctness of two non-trivial algorithms: 
the HHL and Shor's algorithms in Subsections~\ref{sec:hhl} and~\ref{sec:shor}, respectively.

\subsection{HHL Algorithm}\label{sec:hhl}

The HHL algorithm \cite{HHL09} aims to obtain a vector $x$ such that $Ax=b$, where $A$ is a given Hermitian operator and $b$ is a given vector. Suppose that $A$ has the spectral decomposition
$A$=$\Sigma_{j}\lambda_{j}\ket{j}\bra{j}$, where each $\lambda_{j}$ is an eigenvalue and $\ket{j}$ is the corresponding eigenvector of $A$. On the basis %consists of 
$\{\ket{j}\}_{j \in J}$, we have $A^{-1}=\Sigma_{j}\lambda_{j}^{-1}\ket{j}\bra{j}$ and $\ket{b}$=$\Sigma_{j}b_{j}\ket{j}$. Then the vector $x$ can be expressed as 
$\ket{x}$=$A^{-1}\ket{b}=$$\Sigma_{j}\lambda_{j}^{-1}b_{j}\ket{j}$. Here we require that $\ket{j}, \ket{b}$ and $\ket{x}$ are all normalized vectors. Hence, we have
\begin{equation}\label{e:x}
\sum_j|\lambda_j^{-1}b_j|^2=1.
\end{equation}

A quantum program implementing the HHL algorithm is presented in Table~\ref{tab:hhl-alg}.  
The $n$-qubit subsystem $\overline{p}$ is used as a control system in the phase estimation step with $N=2^{n}$. The $m$-qubit subsystem $\overline{q}$ stores the vector $\ket{b}=\sum_{i}\mathit{b}_{i}\ket{i}$. The one-qubit subsystem $r$ is % indicator of 
used to control the while loop. The measurement $\mathit{M}=\{\mathit{M}_{0},\mathit{M}_{1}\}$ in the loop is the simplest two-value measurement: $\mathit{M}_{0}=\ket{0}_{\mathit{r}}\bra{0}$ and $\mathit{M}_{1}=\ket{1}_{\mathit{r}}\bra{1}$. The results of measurements will be assigned  
to the classical variable $\mathit{v}$, which is initialized to be 0. If the value of $\mathit{v}$ is $0$, then  the while loop is repeated until  it is $1$. The unitary operator $\mathit{U}_{b}$ is 
assumed to map $\ket{0}^{\otimes\mathit{m}}$ to $\ket{b}.$
 The controlled unitary operator $\mathit{U}_{f}$ has 
 a control system $\overline{p}$ and a target system $\overline{q}$, that is,

\[
\mathit{U}_{f}=\sum_{\tau=0}^{N-1}\ket{\tau}_{p}\bra{\tau} \otimes U^{\tau},
\]
where $U=e^{iAt}$. Equivalently, we have $U\ket{j}$=$e^{i\lambda_{j}t}\ket{j}$. 
We denote $\phi_{j}$=$\frac{\lambda_{j}t}{2\pi}$ and $\widetilde{\phi_{j}}$=$\phi_{j}$$\cdot N$, then we have $U\ket{j}$=$e^{2\pi i \phi_{j}}\ket{j}$. 
Following~\cite{Zhou23}, to make the algorithm exact, we further assume that there exists $t \in \mathbb{R}$, for all $j \in J$, $\widetilde{\phi}_j=\frac{\lambda_j t \cdot N}{2\pi} \in \{1,2, \cdots, N-1\} $.
  A given controlled unitary operator $\mathit{U}_{c}$ has control system $\overline{p}$ and target system  $r$, more precisely, 
\[
\mathit{U}_{c}\ket{0}_{\overline{p}}\ket{0}_{r}=\ket{0}_{\overline{p}}\ket{0}_{r},\qquad
\mathit{U}_{c}\ket{j}_{\overline{p}}\ket{0}_{r}=\ket{j}_{\overline{p}}(\sqrt{1-\tfrac{C^2}{j^2}}\ket{0}+\tfrac{C}{j}\ket{1})_{r},
\]
 where $1\leq j\leq N-1$ and  $C$ is a
  given non-zero parameter such that $ | \frac{C}{j}| \le 1$ for any $0 \le j  \le N-1$. The symbol $\mathit{H}^{\otimes n}$ represents $n$  Hadamard gates applied to the variables $\overline{p}$; $\mathrm{QFT}$ and $\mathrm{QFT^{-1}}$ are the quantum Fourier transform and the inverse quantum Fourier transform acting on the variable $\overline{p}$. 

\begin{table}[t]
 \caption{A quantum program for the HHL algorithm}
 \label{tab:hhl-alg}
\[
\begin{aligned}
	\mathbf{HHL} &\triangleq\\
	&1: v:=0&&\\
	&2: \mathbf{while}& & v=0 \ \ \mathbf{do}\\
	&3:& &\overline{p}:=\ket{0}^{\otimes n}; \\
	&4:& &\overline{q}:=\ket{0}^{\otimes m}; \\
	&5:& &r:=\ket{0}; \\
	&6: && U_{b}[\overline{q}]; \\
	&7: &&H^{\otimes n}[\overline{p}]; \\
	&8: && U_{f}[\overline{p}\overline{q}]; \\
	&9: && \mathrm{QFT^{-1}}[\overline{p}]; \\
	&10: && U_{c}[\overline{p}r]; \\
	&11:& & \mathrm{QFT}[\overline{p}]; \\
	&12: && U_{f}^{\dagger}[\overline{p}\overline{q}]; \\
	&13: && H^{\otimes n}[\overline{p}]; \\
	&14:& & v:=M[r] \ \mathbf{od}
\end{aligned}
\]
\end{table}

The  correctness of the HHL algorithm can be specified by the Hoare triple:
\[
 \Key{\{\ \mathbf{true}\ \}} \ \mathbf{HHL} \  \Key{\{\ \ket{x}_{\overline{q}}\ \}}\ .
\]
 To prove the correctness, we first consider the loop body of the program in Table~\ref{tab:hhl-alg} and give its proof outline in Table~\ref{t:body1} of Appendix~\ref{appB}.
Now let $D\triangleq  (v=0) \oplus (( \ket{0}^{\otimes n}_{\overline{p}}\ket{x}_{\overline{q}}|1\>_r) \wedge (v=1))$, and $S$ the body of the while loop of the HHL algorithm. From the reasoning in Table~\ref{t:body1}, we see that
\[
\Key{\{\  v=0 \  \}}  \ \mathit{S} \  \Key{\{\ D\ \}} \ .
\]
So $D$ is an invariant of the while loop of the HHL algorithm.  Then by rule [While] we obtain that
\[
\Key{\{\ D\ \}} \ \mathbf{while} \  (\mathit{v}=0) \ \mathbf{do} \  \mathit{S} \ \mathbf{od}\ \Key{\{\ \ket{0}^{\otimes n}_{\overline{p}}\ket{x}_{\overline{q}} \ket{1}_{r} \wedge  (v=1) \ \}} \ .
\]
Finally, we can establish the correctness of the HHL algorithm as given in Table~\ref{tab:hhl-pro}, where we highlight the invariant of the while loop in red. 
\begin{table}[htb]
\caption{Proof outline of the HHL algorithm}
\label{tab:hhl-pro}
\[
\begin{aligned}
  &\Key{\{\ \mathbf{true}\ \}}\\
  &\Key{\{\ (v=0)[0/v]\ \}\qquad\mbox{by rule [Conseq]}}\\
  & v:=0;\\
  &\Key{ \{\ v=0\ \} \qquad\mbox{by rule [Assgn]}}\\
  & \textcolor{red}{\{\  (v=0) \oplus ( \ket{0}^{\otimes n}_{\overline{p}} \ket{x}_{\overline{q}} \ket{1}_{r} \wedge (v=1))\ \} \qquad\mbox{by rule [Oplus]}}\\
 &\mathbf{while} \ v=0 \ \mathbf{do}\\
	&\qquad \Key{\{\  v=0\ \} }\\
	&\qquad \mathit{S}\\
	&\qquad \textcolor{red}{\{\  (v=0) \oplus ( \ket{0}^{\otimes n}_{\overline{p}} \ket{x}_{\overline{q}} \ket{1}_{r} \wedge (v=1))\ \}}\\
 &\mathbf{od}\\
	&\Key{\{\ \ket{0}^{\otimes n}_{\overline{p}} \ket{x}_{\overline{q}} \ket{1}_{r} \wedge  (v=1) \ \} \qquad\mbox{by rule [While]}} \\
  & \Key{\{\  \ket{0}^{\otimes n}_{\overline{p}} \ket{x}_{\overline{q}} \ket{1}_{r} \ \} \qquad\mbox{by rule [Conseq]}}\\
  & \Key{\{\  \ket{0}^{\otimes n}_{\overline{p}}\odot \ket{x}_{\overline{q}}\odot \ket{1}_{r} \ \}\qquad\mbox{by rule [OdotT]}}\\
  & \Key{\{\  \true\odot \ket{x}_{\overline{q}}\odot \true \ \}\qquad\mbox{by rule [PT]}}\\ 
	& \Key{ \{\ \ket{x}_{\overline{q}}\ \} \qquad\mbox{by rule [OdotE]}}
\end{aligned}
\]
\end{table}

	\subsection{Shor's Algorithm}\label{sec:shor}
        Shor's algorithm  relies on the order-finding algorithm \cite{Sho94}. So we  first  verify the correctness  of the latter. Given two co-prime positive integers $x$ and $N$, the smallest positive integer $r$ that satisfies the equation $x^{r}=1(\mathrm{mod})N$ is called the order of $x$ modulo $N$, denoted by $\mbox{ord}(x,N)$.
        The problem of order-finding is to find the order $r$ defined above, which is solved by the program presented in Table~\ref{tab:of-alg}.
Let $ L\triangleq \ulcorner \log(N) \urcorner $, $\epsilon \in (0,1)$  and  $t \triangleq 2L+1+ \ulcorner \log(2+1/2\epsilon ) \urcorner$. The order-finding algorithm can successfully obtain the order of $x$ with probability at least $(1-\epsilon)/2\log(N)$, by using O($L^{3}$) operations as discussed in \cite{NC00}.

\begin{table}
\caption{A quantum program for the order-finding algorithm}
\label{tab:of-alg}
\centering
\begin{align*}
	\mathbf{OF}(x,N):\equiv\\
	1: & z:=1;  \\
	2: & b:=x^{z} \ (\mathrm{mod}\  N); \\
	3:  &\mathbf{while} \ (b\neq 1) \ \mathbf{do}\\
	4: &\qquad \overline{q}:= \ket{0}^{\otimes t};\\
	5:	&\qquad \overline{p}=\ket{0}^{\otimes L};\\
	6: & \qquad H^{\otimes t}[\overline{q}];\\
	7:& \qquad U_{+}[\overline{p}];\\
	8:& \qquad \mathrm{CU}[\overline{q} \overline{p}];\\
	9:& \qquad \mathrm{QFT^{-1}}[\overline{q}];\\
	10:& \qquad z':=M[\overline{q}];\\
	11:& \qquad z:=f(\frac{z'}{2^t});\\
	12:& \qquad b:=x^{z} \ (\mathrm{mod} \ N) \ \mathbf{od}
\end{align*} 
\end{table}

  The variables in $\overline{q}$ correspond to a $t$-qubit subsystem while $\overline{p}$ represents an $L$-qubit subsystem. We introduce the variable $z$ to store the order computed by the program $\mathbf{OF}$, and initialize it to $1$. The unitary operator $\mathit{U}_{+}$  maps $ \ket{0}$ to $\ket{1}$, that is $\mathit{U}_{+}\ket{0}=\ket{1}$. The notation $\mathit{H}^{\otimes t}$ means $t$ Hadamard gates applied to the system $\overline{q}$ and $\mathrm{QFT^{-1}}$ is the inverse quantum Fourier transform. The function $\mathit{f}(x)$  stands for the continued fraction algorithm which returns the minimal denomination $n$ of all convergents $m/n$ of the continued fraction for $x$ with $|m/n-x|<1/(2n^{2})$~\cite{FY21}. The unitary operator $\mathrm{CU}$ is the controlled-$\mathit{U}$, with $\overline{q}$ being the control system and $\overline{p}$  the target system, that is $\mathrm{CU}$ $\ket{i}_{\overline{q}}$$\ket{j}_{\overline{p}}=\ket{i}_{\overline{q}}\mathit{U}^{i}\ket{j}_{\overline{p}}$, where for each $0\leq y \leq 2^{L}$,
  
\[
\begin{aligned}
	U\ket{y}=\begin{cases} 
		\ket{xy\ \mathrm{mod} \ N}  &\ \mathrm{if} \ y\leq N\\
		\ket{y}   \         &\mathrm{otherwise}.
	\end{cases}
\end{aligned}
\]
Note that the states defined by
\[\ket{u_s}\triangleq\frac{1}{\sqrt{r}}\sum_{k=0}^{r-1}e^{-2\pi isk/r}\ket{x^k \mbox{ mod }N}\]
for integer $0\leq s\leq r-1$ are eigenstates of $U$ and
$\frac{1}{\sqrt{r}}\sum_{s=0}^{r-1}\ket{u_s}=1$. Similarly, we assume that for all $0\le s \le r-1$, $(\frac{s}{r}\cdot (2^t)) \in \{0,1,\cdots,2^t-1 \} $.

The variable $b$ stores the value of ($x^{z}$ mod $N$), and the operator ($a$ mod $b$)  computes the modulo of $a$ divided by $b$. If the value of $b$ is not equal to $1$, which means that  the value of $z$ computed by $\mathbf{OF}$ is not equal to the actual order of $x$  modulo $N$, then the program will repeat the body of the while loop until $b=1$. The while loop in the $\mathbf{OF}$ program exhibits probabilistic behaviour due to a measurement in the loop body.

The correctness of the order-finding algorithm can be specified as 
\[
\Key{\{\ 2\leq x \leq N-1 \wedge \gcd(x,N)=1 \ \}} \ \mathbf{OF} \ \Key{\{\ 2\leq x \leq N-1 \wedge z=r\ \}} \ .
\]
To prove this correctness, we first consider the loop body of the order-finding algorithm, as displayed in Table~\ref{t:body2} of Appendix~\ref{appB}.
Now let $\mathit{D'}\triangleq  (z=r \wedge b=1 ) \oplus (z < r \wedge b\neq 1 )$, and $\mathit{S'}$ the body of the while loop $\mathbf{OF}$. From the reasoning in Table~\ref{t:body2}, we specify the correctness of $S'$ as follows:
\[
\Key{\{\ z < r \wedge b \neq 1  \ \}} \ S' \ \Key{\{\ \mathit{D'}\ \}} \ .
\]
So the invariant of the while loop of $\mathbf{OF}$ can be $\mathit{D'}$, and by rule [While] we have 
\[
\Key{\{\ \mathit{D'}\ \}} \ \mathbf{while} \ (b\neq 1) \ \mathbf{do} \ \mathit{S'} \ \mathbf{od} \ \Key{\{\  z=r \wedge b=1 \ \}} \ .
\]
Finally, a proof outline of $\mathbf{OF}$ is given in Table~\ref{tab:of-pro} .

\begin{table}
\caption{Proof outline of the \textbf{OF} program}\label{tab:of-pro}
\begin{align*}
  &\Key{\{ \ 2\le x \le N \wedge \gcd(x,N)=1 \ \}}  \\
  %\gcd(x,N)=1 \wedge N \ \mathrm{mod}\ 2 \ \neq0
  &\In\Key{\{ (z=1)[1/z]\}}  \\
  & z:=1;  \\
  &\Out\Key{\{ z=1\}\quad\mbox{by rule [Assgn]}}  \\
  &\In\Key{\{ (z=1\wedge b=x^1\mbox{ mod }N)[(x^z\mbox{ mod N})/b]\}}  \\
  & b:=x^{z} \ \mathrm{mod} \  N; \\
    &\Out\Key{\{z=1\wedge b=x^1\mbox{ mod }N\}\quad\mbox{by rule [Assgn]}}  \\
	& \In \Key{\{ z < \mbox{ord}(x,N) \wedge b\neq 1  ) \}\quad\mbox{by rule [Conseq]}}  \\
	& \textcolor{red}{ \{ (z=\mbox{ord}(x,N) \wedge b=1 )) \oplus (z < \mbox{ord}(x,N) \wedge b\neq 1 )) \}\quad\mbox{by rule [Conseq]}} \\
	&\mathbf{while} \ (b\neq 1)\  \mathbf{do}\\
	& \qquad \Key{\{z < \mbox{ord}(x,N) \wedge b\neq 1 )\}}\\
	&\qquad \mathit{S'} \\
	& \qquad\textcolor{red}{\{ (z=\mbox{ord}(x,N) \wedge b=1 )) \oplus (z< \mbox{ord}(x,N) \wedge b\neq 1 ))  \}} \\
         & \mathbf{od} \\
	&\Out \Key{\{ z=\mbox{ord}(x,N) \wedge  b=1 )\}\quad\mbox{by rule [While]}}\\
	&\Key{\{ \ 2\leq x \leq N-1  \wedge z=\mbox{ord}(x,N) \ \}\quad\mbox{by rule [Conseq]}}
\end{align*}
\end{table}

Then we introduce Shor's algorithm in Table~\ref{Fig.11}. The function random($a$,$b$) is used to randomly generate a number between $a$ and $b$.
The function gcd($a$,$b$) returns the greatest common divisor  of $a$ and $b$. The operator $\equiv_{N}$  represents identity  modulo $N$. $\mathbf{OF}$($x$,$N$) is the order-finding algorithm given before, which will return the order of $x$  modulo $N$ and assign the order to the classical variable $z$. %Shor's algorithm repeats calling the subroutine $\mathbf{OF}$($x$,$N$) until the order $z$ is even and satisfies $x^{z/2}\equiv_{N}-1$. 
The classical variable $y$  stores one of the divisors of $N$ and we use $y|N$ to represent that $N$ is divisible by $y$.

The correctness of Shor's algorithm can be specified by the Hoare triple:
\[
\Key{\{\ {\rm cmp}(N)\ \}} \ \mathbf{Shor} \ \Key{\{\ y|N \wedge y\neq 1\wedge y\neq N \ \}}
\]
where ${\rm cmp}(N)$ is a predicate stating that $N$ is a composite number greater than $0$. 
%We use the integer $r$ to represent the actual order of $x$  modulo $N$.
The invariant of the while loop in Shor's algorithm  can be 
\[ \begin{aligned}
  & (2\leq x\leq N-1) \wedge ((y= \gcd (x, N) \wedge y\neq N) \\ 
       &  \quad   \qquad \qquad \qquad \quad \vee (y= \gcd (x^{r/2}-1, N) \wedge y\neq N \wedge y \neq 1) \\
     &   \quad  \qquad  \qquad \qquad  \quad  \vee (y= \gcd (x^{r/2}+1, N) \wedge y\neq N \wedge y \neq 1)) . 
     \end{aligned} 
     \]
\begin{table}[t]
\caption{A program for Shor's algorithm}
\label{Fig.11}
\begin{tabular}{ll}
	1: & $\mathbf{if}\ 2 \mid N \ \mathbf{then}$ \\
	2: & \qquad $y:=2$; \\
	3: & $\mathbf{else}$ \\
	4:& \qquad $x:=\mathrm{random}(2,N-1)$; \\
    5:& \qquad $y:=\gcd(x,N)$; \\
	6:& \qquad $\mathbf{while} \ y=1 \ \mathbf{do}$\\
    7:& \qquad \qquad $z:=OF(x,N)$; \\
    8:& \qquad \qquad $\mathbf{if}\ 2\mid z \mbox{ and } x^{z/2}\not\equiv_{N} -1 \ \mathbf{then}$ \\
	% 9:& \qquad \qquad \qquad $ y':=\mathrm{gcd}(x^{z/2}-1,N)$; \\
  9:& \qquad \qquad \qquad $\mathbf{if}\ 1<\mathrm{gcd}(x^{z/2}-1,N)<N \ \mathbf{then}$ \\
    10:& \qquad \qquad \qquad \qquad $y:=\mathrm{gcd}(x^{z/2}-1,N)$; \\
    11:& \qquad \qquad \qquad $\mathbf{else}$ \\
    12:& \qquad \qquad \qquad \qquad $y:=\mathrm{gcd}(x^{z/2}+1,N)$; \\
    13:& \qquad \qquad \qquad $\mathbf{fi}$ \\
    14:& \qquad \qquad $\mathbf{else}$ \\
    15:& \qquad \qquad \qquad $x:=\mathrm{random}(2,N-1)$; \\
    16:& \qquad \qquad \qquad $y:=\gcd(x,N)$; \\
    17:& \qquad \qquad $\mathbf{fi}$ \\
    18:& \qquad $\mathbf{od}$ \\
    19:& $\mathbf{fi}$ \\
\end{tabular}
\end{table}
A proof outline for the correctness of the algorithm is given in Table~\ref{t:shor} of Appendix~\ref{appB}.

\section{Coq Formalization} 

%Analyzing the reasoning process of the algorithms detailed 
As we can see in Section \ref{sec:cases}, %it becomes apparent that 
manual reasoning about the correctness of quantum algorithms are indeed cumbersome. In order to alleviate this burden, 
%the strain of manual reasoning, 
we have formalized all the aforementioned concepts within the Coq proof assistant. Our framework, called CoqQLR, offers several distinct advantages: 1) The majority of the concepts and lemmas presented in this paper have been formalized in Coq with minimal simplification, preserving the full logical expressiveness of the theories; 2) In addition to the rules demonstrated within this paper, we have %established further 
proved some other useful theorems that enhance our capacity for semi-automated reasoning about programs; 3) The algorithms discussed in the last section have all undergone correctness verification within Coq. %eliminating the need for providing extensive preconditions and postconditions.

\begin{figure}[!htbp]
	\centering
	\includegraphics[width=0.7\linewidth]{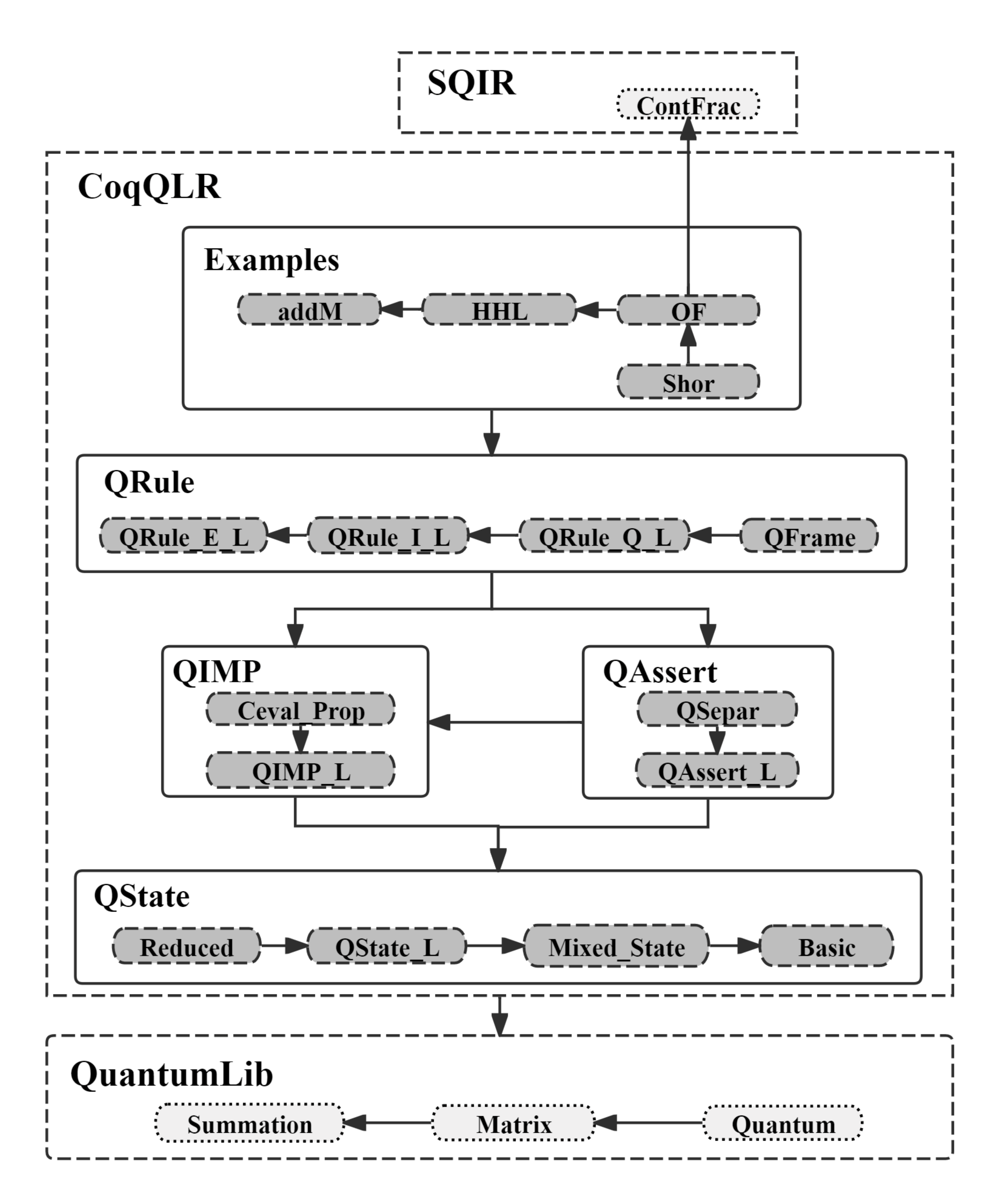}
 \caption{Relationship between different files}
\label{fig:rela}
\Description{Figure 1. Fully described in the text.}
\end{figure}

% \begin{figure}[!ht]
%     \centering
%     \includesvg[width=0.5\textwidth]{Relationship}
%     \caption{Relationship between different files}
%     \label{fig:rela}
% \end{figure}

Our formalization relies on the library \textbf{QuantumLib}\footnote{\url{https://github.com/inQWIRE/QuantumLib.git}}. Figure~\ref{fig:rela} illustrates the interrelationships between different folders. %Upon reviewing the folder names, we can easily identify the particular functions and objectives of each folder. 
In our formalization, we primarily utilize three key files from \textbf{QuantumLib}: \textbf{Summation} for summation operations, \textbf{Matrix} for matrix computations, and \textbf{Quantum} for basic quantum computing concepts.
Building upon the foundation of \textbf{QuantumLib}, we first define the notations of basis vectors and distribution states, along with their associated concepts in the \textbf{QState} folder. The \textbf{QIMP} folder focuses on the syntax and semantics of classical--quantum programs, while the \textbf{QAssert} folder deals with assertions. In the \textbf{QRule} folder, we delineate the formalization of the rules mentioned in Section~\ref{sec:proofsys} and rigorously establish their soundness. Finally, in the \textbf{Examples} folder, we formally prove the correctness of the \textbf{addM} program, HHL algorithm, order-finding algorithm, and Shor's algorithm. 
Below we illustrate the formalization of selected key concepts in Coq. 
More details can be found in our repository of Coq scripts.

In the \textbf{Basic} file, we define the type {\tt Base\_vec} with two parameters $n$ and $i$ of type {\tt nat}, representing the basis vector $\ket{i}$ in an $n$-dimensional space. Some properties of vectors are given in Table~\ref{tab:Base}.
\begin{table}[t]
    \caption{Properties of Basis Vector}
    \label{tab:Base}
    \centering
    \begin{tabular}{c|c}
    \toprule
    Name & Function \\ 
    \midrule
       base1 & $\forall (i \in nat), \ket{i}_{i0}= 1$ \\
        base0 & $\forall (i, j \in nat), \ i \neq j \rightarrow \ket{i}_{j0} = 0$\\
        base\_decom &  $ \forall (n \in nat) (v \in Vector \ n) , 
        v= \sum_i (v_{i0}) \cdot \ket{i}$ \\ 
        base\_inner\_M & $\forall (m, n \in nat) (A:Matrix \ m \ n), 
        \bra{i}A\ket{i}=A_{ii}$\\
        trace\_base &$ \forall (m, n \in nat) (A:Matrix\  m \ n), \sum_i \bra{i}A\ket{i} = \mathrm{tr}(A) $ \\
        base\_inner\_1 & $ \forall (i \in nat), \langle i | i \rangle = 1$ \\
        base\_inner\_0 & $ \forall (i, j \in nat), i \neq j \rightarrow \langle i | j \rangle = 0$ \\
         norm\_base\_1 & $\forall (i \in nat), \Vert \ket{i} \Vert = 1$  \\
         base\_kron & $\forall (i, m, n\in nat), 
         \ket{i/n}^{m} \otimes \ket{i \%n}^{n} = \ket{i}^{m*n}$ \\
        big\_sum\_I & $\sum_i  \ket{i}\bra{i}= I $\\
        qubit0\_base\_kron & $ \forall (i, n \in nat), \ket{0} \otimes \ket{i}^{2^n} = \ket{i}^{2^{(n+1)}}$\\
        qubit1\_base\_kron & $ \forall (i, n \in nat), \ket{1} \otimes \ket{i}^{2^n} = \ket{i+2^n}^{2^{(n+1)}}$\\
        \bottomrule
    \end{tabular}
\end{table}
Before defining the concept of quantum state, we first define  mixed states in the file \textbf{Mixed\_State}. Recall that in \textbf{QuantumLib}, a mixed state is defined as an ensemble with the sum of traces equal to one: 
\begin{verbatim}
Inductive Mixed_State {n} : Matrix n n -> Prop :=
| Pure_S : forall q, Pure_State q -> Mixed_State q
| Mix_S : forall (p : R) q1 q2, 0 < p < 1 -> Mixed_State q1 -> Mixed_State q2 
         -> Mixed_State (p .* q1 .+ (1-p) .* q2), 
\end{verbatim}
where the type {\tt Pure\_State} denotes a pure state.
This definition is equivalent to the notion of density operator. %However, our paper requires the concept of 
Since we mainly work with partial density operators, we define our mixed states by allowing the sum of all traces to be less than one.
\begin{verbatim}
Inductive Mixed_State {n} : Matrix n n -> Prop :=
| Pure_S : forall  q , (0 <= p <= 1) ->  Pure_State q -> Mixed_State (p.* q)  
| Mix_S : forall (p1 p2  : R) q1 q2, 0 <= p1 -> 0 <= p2 -> p1+p2 <= 1 -> Mixed_State q1
          -> Mixed_State q2 -> Mixed_State (p1 .* q1 .+ p2 .* q2).  
\end{verbatim}
When discussing the semantics of programs and the satisfaction relation of assertions, we focus on non-zero states with finite supports. %specifically those quantum states that are non-zero. 
To facilitate the discussion of the properties of these states, we further define the set of non-zero mixed states:
\begin{verbatim}
Inductive NZ_Mixed_State {n} : Matrix n n -> Prop :=
| Pure_S : forall q, (0 < p <= 1) ->  Pure_State q -> NZ_Mixed_State (p .* q)  
| Mix_S : forall (p1 p2  : R) q1  q2, 0 < p1 -> 0 < p2 -> p1+p2 <= 1 -> NZ_Mixed_State q1 
         -> NZ_Mixed_State q2 -> NZ_Mixed_State (p1 .* q1 .+ p2 .* q2).  
\end{verbatim}
\begin{table}[t]
 \caption{Properties of Mixed State}
    \label{tab:mixed-state} 
    \centering
    \begin{tabular}{c|c}
    \toprule
      Name   &  Function    \\
      \midrule 
      pure\_state\_trace\_1 & $\forall \ \rho$,
      Pure\_State $\rho \ \rightarrow $ tr($\rho$) = 1 \\ 
      nz\_mixed\_state\_trace\_real & $\forall \ \rho$,
      NZ\_Mixed\_State $\rho$ $\rightarrow$ Im (tr($\rho$)) = 0\\
      nz\_mixed\_state\_trace\_in01 & $\forall \ \rho$,
      NZ\_Mixed\_State $\rho$ $\rightarrow$  $ 0<\mathrm{Re}((\mathrm{tr}(\rho)))\leq 1$ \\
      nz\_mixed\_state\_Cmod\_1 & $\forall \ \rho$,
      NZ\_Mixed\_State $\rho$ $\rightarrow$  $ 0< |(\mathrm{tr}(\rho))| \leq 1$ \\ 
      nz\_mixed\_state\_Cmod\_plus & $\forall \ \rho_1, \ \rho_2$, NZ\_Mixed\_State $\rho_1$ \\
      &\quad  \qquad $\rightarrow$ NZ\_Mixed\_State $\rho_2$  \\ 
      \ &$\rightarrow$  $ |(\mathrm{tr}(\rho_1 + \rho_2))| = |(\mathrm{tr}(\rho_1))| + |(\mathrm{tr}(\rho_2))|$ \\  
    Cauchy\_Schwartz\_ver1' & $\forall \ (n\in nat) (u, v \in Vector\ n )$, $u \neq \mathbf{0}$ \\ 
    & 
    $\rightarrow  (\neg (\exists (c\in  \mathbb{C}), c * u = v)) $  \\
    & $ \rightarrow |\langle u, v \rangle|^2 <  |\langle u, u \rangle| * |\langle v, v \rangle| $ \\
    mixed\_mult\_trace\_le\_1 & $\forall \ \rho_1, \ \rho_2$, Mixed\_State $\rho_1$ $\rightarrow$ Mixed\_State $\rho_2$  \\ & $\rightarrow$ $0 \leq |(\mathrm{tr}(\rho_1 \times \rho_2))| \leq 1$  \\ 
    nz\_mixed\_mult\_trace\_lt\_1 &  $\forall \ \rho_1, \ \rho_2$, NZ\_Mixed\_State $\rho_1$ $\rightarrow$ NZ\_Mixed\_State $\rho_2$  \\ 
    &$\rightarrow$ $ \neg (\exists \ c \in \mathbb{C}, c * \rho_1 = \rho_2) \rightarrow |(\mathrm{tr}(\rho_1 \times \rho_2))| < 1$  \\ 
    pure\_sqrt\_trace & $\forall \ \rho$, Pure\_State $\rho$ $\rightarrow$ $| \mathrm{trace}(\rho \times \rho)| = 1 $\\ 
    nz\_mixed\_sqrt\_trace & $\forall \ \rho$, 
    NZ\_Mixed\_State $\rho \rightarrow$ $\neg$ Pure\_State $\rho$ \\ 
    & $\rightarrow |
    \mathrm{tr}(\rho \times \rho)|  < 1 $ \\ 
    nz\_mixed\_pure & $\forall \ \rho_1, \ \rho_2, \ \phi$, NZ\_Mixed\_State $\rho_1$ \\
    & \qquad \qquad \  $\rightarrow $ NZ\_Mixed\_State $\rho_2$  \\ & $\rightarrow $ $\Vert \phi \Vert =1$ $\rightarrow$ $\rho_1 + \rho_2 = \ket{\phi}\bra{\phi}$ \\ & $\rightarrow$ 
    $\exists \ p_i, 0<p_i \leq 1 \wedge \rho_i = p_i * \ket{\phi}\bra{\phi}\ \mathrm{for} \ i=1,2$ \\
    \bottomrule
    \end{tabular}
\end{table}
It can be verified that the type {\tt NZ\_Mixed\_State} is preserved after many operations. Additionally, some important properties are summarized in Table~\ref{tab:mixed-state}.

In the \textbf{QState\_L} file, we define classical states, quantum states, and distribution states. Within our framework CoqQLR, we use subscripts to represent variables. For instance, we denote the classical variable \( x_i \) with \( i \) and the set of variables \( \langle q_s, q_{s+1}, \ldots, q_{e-1} \rangle \) with \( (s, e) \). We define the type \texttt{cstate} as a Coq type of \texttt{list nat}, representing a classical state where the \( i \)-th element corresponds to the value of variable \( x_i \). Syntactically, we represent a quantum state \( \rho \) with domain \( (s, e) \) using a \( 2^{(e-s)} \)-dimensional matrix. 
\begin{verbatim}
            Definition qstate (s e :nat):= Density (2 ^ (e-s)).
\end{verbatim}
In order to ensure that this definition is well-formed, we require that the matrix represents a mixed state. Similarly, since we are almost exclusively concerned with the properties of non-zero quantum states, we define a type \texttt{WF\_qstate} to further ensure that this matrix is a non-zero mixed state and the domain is valid, that is, \( s \leq e \):
\begin{verbatim}
              Definition WF_qstate {s e : nat} (rho : qstate s e ) :=
              @NZ_Mixed_State (2 ^ (e - s)) rho /\ (s <= e).
\end{verbatim}
Then a state is a pair with product type \texttt{(cstate * (qstate s e))}. A state is considered well-formed if its quantum component is well-formed.

Next, we proceed to define the notion  of  distribution state. Notably, a distribution is characterized by a mapping with a finite support set. To represent this distribution, we utilize the \texttt{FMapList} structure in Coq, which employs a \texttt{list} to encapsulate the finite mapping. A finite mapping from type {\tt X.t} to type {\tt elt} can be  created with the core \texttt{FMapList.Make(X).t(elt)}, with the following internal structure:
\begin{verbatim}
Record slist (elt:Type) := 
{this :> list (X.t * elt); sorted: sort (KeyOrderedType X.ltk elt) this}.
Definition t (elt:Type): Type := slist elt.
\end{verbatim}
The structure is composed of two parts: \texttt{this} and \texttt{sorted}. The \texttt{this} component uses a list to represent a finite mapping \( f \), where each element is a pair \((a, b)\) of types \texttt{(X.t * elt)}, indicating that \( f(a) = b \). The \texttt{sorted} aspect ensures that the list is sorted, which is essential for maintaining distinct domains and ensuring that functions \( f_1 \) and \( f_2 \), defined by the structure \texttt{FMapList}, are equal if and only if their list components are identical. This sorted requirement implies that type \texttt{X} must be an \texttt{OrderedType}. 
Consequently, we define the type {\tt dstate} to represent the distribution state by the following code, a mapping from {\tt cstate} to {\tt qstate}:
\begin{verbatim}
           Module Import StateMap := FMapList.Make(Cstate_as_OT).
           Definition dstate (s e:nat) := StateMap.t (qstate s e).
\end{verbatim}
The module \texttt{Cstate\_as\_OT} defines \texttt{cstate} as an \texttt{OrderedType}. We employ an element of type \texttt{dstate} to document the mapping relationships within the support set of a distribution \(\mu\), thereby any element absent from its list part has the default value \(\mathbf{0}\). 
\begin{verbatim}
Definition option_qstate {s e:nat} (q: option (qstate s e)): (qstate s e):=
 match q with 
  |None => Zero
  |Some  x => x
 end.
Definition d_find {s e:nat} (sigma:cstate) (mu: dstate s e): qstate s e := 
option_qstate (StateMap.find sigma mu).
\end{verbatim}
To guarantee the well-formedness of this construct, we introduce the type \texttt{WF\_dstate}.
\begin{verbatim}
Inductive WF_dstate_aux {s e:nat}: list(cstate * (qstate s e)) -> Prop:= 
|WF_nil: WF_dstate_aux nil
|WF_cons st mu': WF_state st -> WF_dstate_aux mu' -> (d_trace_aux (st::mu'))<= 1 
                 -> WF_dstate_aux (st::mu').
Definition WF_dstate {s e:nat} (mu: dstate s e):Prop:= WF_dstate_aux (this mu).
\end{verbatim}
This type is defined with the requirement that any quantum state mapped by an element within the support set is a non-zero mixed state, and the sum of their traces must be less than 1.
Naturally, we have proved in Coq that the well-formedness of quantum states, states, and distribution states are preserved under their respective operations.
In addition, we present some useful properties in Table~\ref{tab:QState}.
\begin{table}[t]
 \caption{Properties of Distribution States}
    \label{tab:QState}
    \centering
 \begin{tabular}{c|c}
 \toprule
       Name  & Function  \\
       \midrule
       d\_trace\_scale & $\forall \ \mu, \ p$, $ 0 \le p \rightarrow \Vert p * \mu  \Vert 
       = p * \Vert \mu \Vert $ \\
       d\_app\_empty\_l & $\forall \ \mu$,
       $\mu + \varepsilon = \mu $ \\ 
       d\_trace\_app & $\forall \ \mu_1, \ \mu_2$, $\Vert \mu_1+\mu_2 \Vert = \Vert \mu_1 \Vert + \Vert \mu_2 \Vert$ \\ 
       d\_trace\_le\_1\_big\_dapp & $\forall 
       \ \{p_i\}_{i \in I}, \ \{\mu_i\}_{i \in I}, 
       (\forall \ i, 0 < p_i) \rightarrow \ \Vert \sum_{i} p_i \mu_i \Vert \leq \sum_{i} p_i$ \\
       d\_find\_empty & $\forall \  \sigma, \ \varepsilon(\sigma)=\mathbf{0} $\\ 
       d\_find\_scale & $\forall \ \mu, \ p,  \  \sigma$,
       $(p * \mu) (\sigma) = p * (\mu(\sigma))$ \\
       d\_find\_app & $\forall \ \mu_1, \  \mu_2, \ \sigma$, $(\mu_1 + \mu_2)(\sigma) = \mu_1(\sigma_1) + \mu_2 (\sigma)$ \\ 
      d\_scale\_empty & $\forall \ p, \mu$, 
       $p * \varepsilon = \varepsilon$ \\
d\_scale\_integral & $\forall \ p, \ \mu$, $p * \mu =\varepsilon \rightarrow p=0 \vee \mu = \varepsilon$\\
       d\_scale\_1\_l & $\forall \ \mu$, 
       $1 * \mu = \mu$ \\
       d\_scale\_assoc & $\forall \ p_1, \ p_2, \mu$, $p_1 * (p_2 * \mu)=(p_1*p_2) * \mu$ \\
       d\_app\_comm & $\forall \ \mu_1, \ \mu_2$, $\mu_1+\mu_2= \mu_2+\mu_1$\\
       d\_app\_assoc & $\forall \ \mu_1, \mu_2, \ \mu_3$, $\mu_1+\mu_2 +\mu_3= \mu_1+(\mu_2+\mu_3)$\\ 
       d\_scale\_app\_distr & $\forall \ p, \ \mu_1,\ \mu_2$, $p * (\mu_1+\mu_2) = p * \mu_1 + p *\mu_2$ \\
       dstate\_equal & $\forall \ \mu_1, \ \mu_2$, $\mu_1=\mu_2 \leftrightarrow (\forall \ \sigma), \mu_1(\sigma)=\mu_2(\sigma)$\\
       \bottomrule
    \end{tabular}
\end{table}

In the \textbf{Reduced} file, we formalize the notation $\rho|_{V}$ using the \texttt{Reduced} function and $\mu|_V$ using the \texttt{d\_reduced} function. Similarly, we employ the subscript pair $(l,r)$ to denote the set of quantum variables $V = \langle q_l, q_{l+1}, \cdots, q_{r-1} \rangle$.
A selection of lemmas and their corresponding functions are listed in Table~\ref{tab:Reduced}. The table includes examples for \texttt{Reduced}, while properties for \texttt{d\_reduced} are analogous and thus omitted for brevity.

\begin{table}[t]
   \caption{Properties of Restriction}
    \label{tab:Reduced}
    \centering
    \begin{tabular}{c|c}
    \toprule
        Name &  Function  \\ 
        \midrule
        Reduced\_trace &  $\forall \ \rho, \ V. \ \mathrm{tr} (\rho|_{V} ) = \mathrm{tr} (\rho)$  \\ 
        Reduce\_Zero &  $ \forall \ V. \ \mathrm{tr} (\mathbf{0}|_{V}) = \mathbf{0} $   \\ 
         Reduced\_refl &  $\forall \ \rho, \ \rho|_{\mathrm{dom}(\rho)}= \rho$  \\ 
        Reduced\_tensor\_l &  $ \forall \ \rho_1, \rho_2. \ (\rho_1 \otimes \rho_2)|_{\mathrm{dom}(\rho_1)} = \mathrm{tr}(\rho_2)*\rho_1 $  \\ 
        Reduced\_tensor\_r &  $ \forall \ \rho_1, \rho_2. \ (\rho_1 \otimes \rho_2)|_{\mathrm{dom}(\rho_2)} = \mathrm{tr}(\rho_1)*\rho_2 $  \\
        Reduced\_scale &  $ \forall \ \rho, V, p. \  (p * \rho)|_{V} = p * (\rho|_{V})$  \\ 
        Reduced\_plus & $ \forall \ \rho_1, \rho_2, V. \  (\rho_1 + \rho_2)|_{V} = (\rho_1|_{V}) + (\rho_2|_{V})$  \\ 
        big\_sum\_Reduced & $\forall \ \{p_i\}_{i \in I}, V.  \  (\sum_i (\rho_i))|_{V} = \sum_i ((\rho_i)|_{V})$     \\ 
        Reduced\_assoc &  $ \forall \ \rho, \ V_1, V_2, V_1 \subseteq V_2 \rightarrow (\rho|_{V_2})|_{V_1} = \rho |_{V_1} $      \\ 
        % Reduced-comm &    \\ 
        \bottomrule
    \end{tabular}
\end{table}

The formalization of our classical--quantum languages, assertions, and rules follows the structure of classical Hoare Logic. The syntax and semantics of our programming language are defined in the \textbf{QIMP\_L} file. The syntax of commands is specified by the type {\tt com}. Let \(\mu\) be of type \texttt{dstate}. The semantics of a program \(c\) applied to \(\mu\) are determined by the function \texttt{ceval} in Coq, which applies \(c\) to each element in the list component of \(\mu\). 
\begin{verbatim}
Inductive ceval {s e:nat}: com -> dstate s e-> dstate s e-> Prop:=
|E_com:  forall c (mu mu':dstate s e), WF_dstate mu 
-> (ceval_single c (StateMap.this mu) (StateMap.this mu')) -> ceval c mu mu'.
\end{verbatim}
The \texttt{ceval\_single} function is sufficiently complex that we omit it here. For the sake of simplification, we formalize the semantics of the \(\mathbf{while}\)-statement as:
\[
\begin{aligned}
\denote{\While b \Then c \ \Od}_{(\state,\qstate)}  & = & \left\{\begin{array}{ll}
 \denote{c; \ \While b \Then c \ \Od}_{(\state,\qstate)}  & \mbox{if $\denote{b}_\state=\true$}\\
 \denote{\Skip}_{(\state,\qstate)}  & \mbox{if $\denote{b}_\state=\false$} . 
        \end{array}\right.
\end{aligned}
\]
This formalization is equivalent to the definition provided in our paper, as discussed in detail in the work of~\cite{DENG202273}. Additionally, we consider only those statements that terminate successfully.
Furthermore, to ensure that the definition of \texttt{ceval} is well-defined, we establish two lemmas: \texttt{ceval\_sorted} and \texttt{WF\_ceval}. 

\begin{verbatim}
Lemma ceval_sorted {s e:nat}: forall c (mu mu':list (cstate *qstate s e)) 
(Hm: Sorted (StateMap.Raw.PX.ltk (elt:=qstate s e)) mu)
(H:ceval_single c mu mu'),
Sorted (StateMap.Raw.PX.ltk (elt:=qstate s e)) mu'.
\end{verbatim}

\begin{verbatim}
Lemma WF_ceval {s' e':nat}: forall c (mu mu':list (cstate * qstate s' e')),
WF_dstate_aux mu -> ceval_single c mu mu' -> WF_dstate_aux mu'. 
\end{verbatim}
The former ensures the sorted order of the output list, enabling the output distribution to be well-defined as a \texttt{dstate}, while the latter guarantees that the well-formedness of \texttt{dstate} is preserved, demonstrating that each statement that terminates successfully maintains the trace. 

The \textbf{Ceval\_Prop} file contains two key theorems: {\tt ceval\_big\_dapp} and {\tt Reduced\_ceval\_swap}. 
The {\tt ceval\_big\_dapp} theorem characterizes the linear behavior of the function \([\![c]\!]\), considering those coefficients \( p_i \) such that \( p_i > 0 \).
The {\tt Reduced\_ceval\_swap} theorem elucidates the commutativity of the function with the restriction of the distribution.

The \textbf{QAssert\_L}  file introduces the syntax and semantics of the assertion language as defined in Section~\ref{sec:ass}. 
%Their formalizations are recursively defined in Coq, following the recursive definitions presented in Section~\ref{sec:proof}.
Some useful properties such as Lemma~\ref{lem:sat_linear} and~\ref{lem:sat_wedge_odot}
are proved  within this document. Furthermore, Lemma~\ref{lem:sat_res}, \ref{lem:seman-odot-eq} and \ref{lem:seman-separ} are shown in the file titled \textbf{QSepar}. It is worth mentioning that in the \textbf{QSepar} file, 
we focus exclusively on assertions for which the sets of quantum variables can be expressed as pair $(s, e)$, as specified by {\tt Considered\_Formula}.

The rules in Table~\ref{tab:rule_I}, except for the [QFrame] rule, are validated in the \textbf{QRule\_I\_L} file. The [QFrame] rule is specifically addressed in the \textbf{QFrame} file. The key to proving this rule is based on Lemma~\ref{lem:sat-pre}. Similarly, we restrict our attention to the assertions that  satisfy {\tt Considered\_Formula} in this file.
% where the quantum variable sets are representable as the pairs in the form $(s, e)$. 
The rules in Table~\ref{tab:rule_Q} are shown to be sound in \textbf{QRule\_Q\_L}. The validity of the rules presented in Table~\ref{tab:rule_E} is established within the context of \textbf{QRule\_E\_L}.
In addition to the rules mentioned above, we also demonstrate some useful lemmas that aid in better automated reasoning about algorithms. All the rules are formalized as theorems whose names are prefixed with "rule\_". Let us consider the [QInit] rule as an example:

\begin{verbatim}
                    Theorem rule_QInit: forall s e,
                    {{BTrue}}   
                    <{ [[ s e ]] : Q= 0 }> 
                    {{(QExp_s s e (| 0 > _ (2 ^ (e-s))))}}.
                    Proof. 
                    unfold hoare_triple. 
                    intros s e s' e' (mu,IHmu) (mu', IHmu'). 
                    intros.   
                    inversion_clear H; simpl in H2.  
                    rewrite sat_Assert_to_State in *.  
                    inversion_clear H0.   
                    apply sat_F. 
                    eapply WF_ceval. apply H. apply H2. 
                    apply rule_Qinit_aux' with  mu.  
                    intuition. intuition. assumption.  
                    Qed.
\end{verbatim}

Finally, in the \textbf{Examples} folder, we demonstrate the correctness the algorithms presented in this paper. In the verification of the order-finding algorithm, we referred to a file named \textbf{ContFrac}, which involves the continued fraction algorithm and has been formally established by the work of~\cite{peng2023formally}. 
Let us use  the \textbf{addM} program as a case study to elucidate the process of algorithmic verification in Coq, by employing our developed framework. That program is formalized as follows:

\begin{verbatim}
                            Definition addM : com := 
                          <{ [[0 1]] :Q= 0 ;
                             [[1 2]] :Q= 0 ;
                             hadamard [[0 1]];
                             hadamard [[1 2]];
                             v1 :=M [[0 1]];
                             v2 :=M [[1 2]];
                             v := (AId v1) + (AId v2) }>.
\end{verbatim}
The correctness of the above program can be established by the following theorem.

\begin{verbatim}
Lemma correctness_addM:  
{{ BTrue }}
addM 
{{APro [((1/2), (SPure (BEq (AId v) 1))) ; ((1/2), SPure (BNeq (AId v) 1))]}}.
Proof. 
(*The proof of the initialization part*)
eapply rule_seq. eapply rule_QInit. simpl sub.  
eapply rule_seq. eapply rule_conseq_l. apply rule_OdotE.
eapply rule_qframe' with (F2:= | | 0 ⟩_ (2) >[ 1, 2]); simpl; try lia. 
apply Qsys_inter_empty'; try lia.
split. eapply rule_QInit. split. apply inter_empty. left. reflexivity. lia. 
...
Qed.
\end{verbatim}

In summary, our Coq development contains over 29,000 lines of code. A detailed view of the code statistics for each file is given in Table~\ref{tab:cm}.

\begin{table}[t]
    \caption{Code Metric}
    \label{tab:cm}
    \centering
    \begin{tabular}{ c| c| c| c  }
    \toprule
    \ Documents \ & \ Files \  & \ Lines of Code  \ & \ Total  \ \\ 
    \midrule 
   QState &  Basic    &  600 & 5700 \\ 
          & Mixed\_State & 2000 &   \\ 
          & QState\_L & 2200 &    \\ 
          & Reduced & 900  &   \\  
    \bottomrule
    QIMP & QIMP\_L & 2200 & 4100 \\
        &  Ceval\_Prop & 1900 &  \\ 
       \hline 
   QAssert &  QAssert\_L & 2000&  7000  \\ 
           & QSepar & 5000 &   \\ 
           \hline 
    QRule & QRule\_E\_L & 2500 & 8500 \\ 
               & QRule\_I\_L & 2300  &   \\ 
               & QRule\_Q\_L & 1300 &  \\
               & QFrame & 2400 &  \\
               \hline 
      Examples &  addM & 500 &   3700 \\
               & HHL & 1100 & \\
                & OF & 1600 &  \\
                & Shor & 500  &   \\ 
    \hline 
    Total &    &     &   29000  \\ 
    \hline 
    \end{tabular}
\end{table}

\label{sec:imple}

\section{Conclusion and Future Work}\label{sec:con}
We have presented a new quantum Hoare logic for classical--quantum programs. It includes distribution formulas for specifying probabilistic properties of classical assertions naturally, and at the same time allows for local reasoning. We have proved the soundness of the logic with respect to a denotational semantics and exhibited its usefulness in reasoning about the functional correctness of  the HHL and Shor's algorithms, which are  non-trivial algorithms involving probabilistic behaviour due to quantum measurements and unbounded while loops. Finally, we have embedded the logic into the Coq proof assistant to alleviate the burden of manually reasoning about classical--quantum programs.

 We have not yet precisely delimited the expressiveness of our logic. It is unclear whether the logic is relatively complete, which is an interesting future work to consider.
% We would also like to embed the logic into a proof assistant so to alleviate the burden of manually reasoning about quantum programs as done in Section~\ref{sec:cases}.
Usually there are two categories of program logics when dealing with probabilistic behaviour:  satisfaction-based or expectation-based \cite{DENG202273}. Our logic belongs to the first category. In an expectation-based logic, e.g. the logic in  \cite{Yin12,Yin16}, the Hoare triple $\{P\} c \{Q\}$ is valid in the sense that the expectation of an initial state satisfying $P$ is a lower bound of the expectation of the final state satisfying $Q$. 
It would be interesting to explore local reasoning in expectation-based logics for classical--quantum programs such as those proposed in \cite{FY21,FLY22}. 

\begin{acks}
We would like to thank Professor Jean-François Monin  for his valuable suggestions on the implementation of Coq. The work was supported by the National Key R\&D Program of China under Grant No.
2023YFA1009403, the National Natural Science Foundation of China under Grant Nos. 62472175, 62072176, and 12271172, Shanghai Trusted Industry Internet Software Collaborative Innovation Center, and the ``Digital Silk Road'' Shanghai International Joint Lab of Trustworthy Intelligent Software under Grant No. 22510750100.
\end{acks}
\bibliographystyle{ACM-Reference-Format}
\bibliography{ref}
\appendix

\section{Proof of Proposition~\ref{prop:closed}}
\label{sec:closed}

Before delving into the proof of Proposition~\ref{prop:closed}, we establish some technical lemmas that will be instrumental in our argument.

\begin{lemma}
\label{lem:seman-separ}
Assume that \(V \subseteq \mathbf{QVar}\). Let \(\sigma\) be a classical state, \(\rho \in \mathcal{D}^{-}(\mathcal{H}_{V})\) a quantum state, and \(F\) an assertion. If \((\sigma, \rho) \models F\), there exist \(\rho_1 \in \mathcal{D}^{-}(\mathcal{H}_{\mathrm{qfree}(F)})\) and \(\rho_2 \in \mathcal{D}^{-}(\mathcal{H}_{V \setminus \mathrm{qfree}(F)})\) such that \(\rho = \rho_1 \otimes \rho_2\).
\end{lemma}

The proofs of Lemma ~\ref{lem:seman-separ} are detailed in our CoqQLR framework.

\begin{lemma}
\label{lem:quan_closed}
Let \( V \subseteq \mathbf{QVar} \) and \(\ket{s}\) be a quantum expression. The set \( \{ \rho \in \padist{\CH_V} \mid \rho \models \ket{s} \} \cup \{\mathbf{0}\} \) is a closed set.
\end{lemma}

\begin{proof}
By Lemma~\ref{lem:seman-separ}, for any \(\rho\), if \(\rho \models \ket{s}\) then there exist $\rho_1 \in \padist{\CH_{\mathrm{qfree}(\ket{s})}}$ and $\rho_2 \in \padist{\CH_{V \backslash \mathrm{qfree}(\ket{s})}}$ such that 
\(
\rho = \rho_1 \otimes \rho_2.
\)
Furthermore, by the semantics of quantum expressions, we have \(\rho|_{\mathrm{qfree}(\ket{s})} = p \cdot \ket{s}\bra{s}\) for some \(p \in (0,1]\). Consequently,
\[
\begin{aligned}
& \{ \rho \in \padist{\CH_V} \mid \rho \models \ket{s} \} \cup \{\mathbf{0}\} \\
= & \{ (p \cdot \ket{s}\bra{s}) \otimes \rho' \mid p \in (0,1], \rho' \in \padist{\CH_{V \backslash \mathrm{qfree}(\ket{s})}} \backslash \{\mathbf{0} \} 
\} \cup \{\mathbf{0}\} \\
= & \{ (p \cdot \ket{s}\bra{s}) \otimes \rho' \mid p \in [0,1], \rho' \in \padist{\CH_{V \backslash \mathrm{qfree}(\ket{s})}} \} \\
= & \{ (p \cdot \ket{s}\bra{s}) \mid p \in [0,1] \} \otimes \padist{\CH_{V \backslash \mathrm{qfree}(\ket{s})}}.
\end{aligned}
\] 
Since the state space \(\{ (p \cdot \ket{s}\bra{s}) \mid p \in [0,1] \}\) and \(\padist{\CH_{V \backslash \mathrm{qfree}(\ket{s})}}\) are both closed sets, it follows that \(\{ \rho \in \padist{\CH_V} \mid \rho \models \ket{s} \} \cup \{\mathbf{0}\}\) is also a closed set.
\end{proof}

\begin{lemma}
   \label{combin}
    Let $A$ and $B$ be two closed sets of distributions, $\lambda$ a real number in $[0,1]$. The set $C = \{ \lambda a + (1-\lambda) b \,|\, a \in A, b\in B\}$ is also closed. 
\end{lemma}

\begin{proof}
    To prove that $C$ is closed, we first observe that for any continuous function $F$, if $a$ and $b$ both range over closed sets, then the range of $F(a,b)$ is also closed. 
   Consider the function \(F(a, b) = \lambda a + (1-\lambda) b\), where \(\lambda\) is a fixed real number in \([0,1]\). It is evident that \(F\) is a continuous function, as it represents a linear combination of \(a\) and \(b\) with fixed coefficients. Therefore, given that \( A \) and \( B \) are closed sets, the set \( C = \{ F (a,b) \,|\, a \in A, b\in B\} \) is closed.
\end{proof}
% \begin{proof}

%    Let $\{c_n\}$ be a sequence in the set $C$, such that 
%     $ c_n= \lambda a_n + (1-\lambda) b_n$ for some $ a_n \in A $ and $b_n \in B$. We aim to show that $ \lim_{n \to \infty} c_n \in C$. 
    
%     Since $A$ and $B$ are both closed, there exist some $a \in A$ and $b \in B$ such that:
%     $$\lim_{n \to \infty} a_n = a \quad \text{and} \quad \lim_{n \to \infty} b_n = b. $$
%     Therefore, we have
%  % \begin{equation}
%  $$
%  \begin{aligned}
%         \lim_{n \to \infty} c_n &= \lim_{n \to \infty} (\lambda a_n + (1-\lambda) b_n) \\
%         &= \lambda \lim_{n \to \infty} a_n + (1-\lambda) \lim_{n \to \infty} b_n \\
%         &= \lambda a + (1-\lambda) b.
%     \end{aligned}
%  % \end{equation}  
% $$

%     Since $a \in A$ and $b \in B$, it immediately follows that $(\lambda a + (1-\lambda) b)\in C $. Thus the set $C$ is closed.
% \end{proof}

\begin{lemma}
\label{Con}
    Let $A$ and $B$ be two closed sets. The set $C = \{ \lambda a + (1-\lambda) b \,| \, a \in A, b\in B, \lambda \in [0,1]\}$ is also closed. 
\end{lemma}

\begin{proof}
The proof follows a similar structure to that of the preceding lemma. Let $F(a, b,\lambda)=\lambda a + (1-\lambda) b$. We get that $F$ is a continuous function. Given that $A$ and $B$ are both closed sets, and noting that the interval [0,1] is also a closed set, it follows that $C=\{F(a,b, \lambda) \ | \  a \in A, b\in B, \lambda \in [0,1]\} $ is a closed set.
\end{proof}

% \begin{proof}
%     Let $\{c_n\}$ be a sequence in the set $C$, such that 
%     $ c_n= \lambda_n a_n + (1-\lambda_n) b_n$ for some $ a_n \in A $, $b_n \in B$ and
%     $\lambda_{n} \in [0,1] $. We aim to show that $ \lim_{n \to \infty} c_n \in C$. 

%     Since $A$ and $B$ are both closed, there exist some $a \in A$ and $b \in B$ such that:  
%     \[ \lim_{n \to \infty} a_n = a \quad \text{and} \quad \lim_{n \to \infty} b_n = b. \]
   
%     Since $\lambda_n$ is bounded,  there exists a 
%     sub-sequence $\lambda_{n_k}$ and a real number $\lambda \in [0,1]$ such that $$\lim_{ {n_k} \to \infty} \lambda_{n_k}=\lambda.
%     $$

%     We now consider the sub-sequence $a_{n_k}$ of $a_n$, $b_{n_k}$ of $b_n$ and $c_{n_k}$ of $c_n$. Note that any sub-sequence of a convergent sequence converges  to the same limit. We have $$ \lim_{n_k \to \infty} a_{n_k} = a, \quad
%     \lim_{n_k \to \infty} a_{n_k} = b, 
%     \quad 
%     \mathrm{and}  \ \lim_{n_k \to \infty} c_{n_k} = c.$$
%     So we have 
%     $$
%     \begin{aligned}
%          \lim_{n_k \to \infty} c_{n_k}
%         &=\lim_{n_k \to \infty} (\lambda_{n_k} a_{n_k} + (1-\lambda_{n_k}) b_{n_k})\\
%         &=\lambda \lim_{n_k \to \infty} a_{n_k}  +  (1-\lambda ) \lim_{n_k \to \infty} b_{n_k}\\
%         &= \lambda a  +  (1-\lambda ) b.
%     \end{aligned}
%  $$
%     Since $a \in A$ and $b \in B$, by the definition of $C$ we see that $(\lambda a  +  (1-\lambda ) b) \in C$. So the set $C$ is closed.
% \end{proof}

% \setcounter{proposition}{0}
\begin{lemma}
\label{lem:State_closed}
Let $V \subseteq \Qvar$. 
For any state formula \( F \),  the set \( [\![F]\!]_{V} \triangleq \{ \mu \mid \mu \models F, \mathrm{dom}(\mu)=V  \}  \) is a closed set.
\end{lemma}

\begin{proof}
We proceed by induction on the structure of \( F \). For brevity, we omit the subscript.

\begin{itemize}
    \item \textbf{Base Case:} \( F \equiv P \). Suppose there exists a sequence of distributions \( \{\mu_n\}_{n \geq 0} \) such that each \( \mu_n \in [\![P]\!] \). Let \( \mu \) be the limit \( \lim_{n \to \infty} \mu_n \) if it exists. We want to show that \( \mu \in [\![P]\!] \). By definition, we need to prove that for any \( \sigma \in \Sigma \), if \( \text{tr}(\mu(\sigma)) > 0 \), then \( \sigma \models P \). Since \( \mu(\sigma) = \lim_{n \to \infty} (\mu_n(\sigma)) \), \( \text{tr}(\mu(\sigma)) > 0 \) implies that there exists \( N \) such that for all \( n > N \), \( \text{tr}(\mu_n(\sigma)) > 0 \). By definition, this implies \( \sigma \models P \).
    
    \item \textbf{Base Case:} \( F \equiv \ket{s} \). Similarly, 
    let \( \{\mu_n\}_{n \geq 0} \) be a sequence of distributions  such that each \( \mu_n \in [\![\ket{s}]\!] \), and let \( \mu \) be its limit. 
    We show that \( \mu \in [\![\ket{s}]\!] \). By definition, we need to prove that for any \( \sigma \in \Sigma \), if \( \text{tr}(\mu(\sigma)) > 0 \), then \( \mu(\sigma) \models \ket{s} \). For any \( \sigma \in \Sigma\), since \( \mu(\sigma) = \lim_{n \to \infty} (\mu_n(\sigma)) \) and \( \mu_n(\sigma) \in \{ \rho \in \padist{\CH_V} \mid \rho \models \ket{s} \} \cup \{\mathbf{0}\}  \) for any \( n\), a closed set by Lemma~\ref{lem:quan_closed},
    it follows that \( \mu(\sigma) \in \{ \rho \in \padist{\CH_V} \mid \rho \models \ket{s} \} \cup \{\mathbf{0}\}  \). If \( \text{tr}(\mu(\sigma)) > 0 \), then \( \mu(\sigma) \neq \mathbf{0} \), meaning that \( \mu(\sigma) \models \ket{s} \).
    
    \item \textbf{Inductive Step:} \( F \equiv F_1 \wedge F_2 \). By Lemma~\ref{lem:sat_wedge_odot}, we see that
     \( [\![F_1 \wedge F_2]\!] = [\![F_1]\!] \cap [\![F_2]\!] \). By the induction hypothesis, both \( [\![F_1]\!] \) and \( [\![F_2]\!] \) are closed sets. Since the intersection of two closed sets is a closed set, we conclude that \( [\![F_1 \wedge F_2]\!] \) is closed.
    
    \item \textbf{Inductive Step:} \( F \equiv F_1 \odot F_2 \). When \( \text{dom}(F_1) \cap \text{dom}(F_2) \neq \emptyset \), we have \( [\![F_1 \odot F_2]\!] = \emptyset \), which is a closed set. Otherwise, we have \( [\![F_1 \odot F_2]\!] = [\![F_1]\!] \cap [\![F_2]\!] \) by Lemma~\ref{lem:sat_wedge_odot}. By the conclusion in the previous item, we see that \( [\![F_1 \odot F_2]\!] \) is also closed.  \qedhere
\end{itemize}
\end{proof}

\begin{proposition}[Proposition~\ref{prop:closed}]
Let $V \subseteq \Qvar$. 
For any assertion \( D \),  the set \( [\![D]\!]_{V} \triangleq \{ \mu \mid \mu \models D, \mathrm{dom}(\mu)=V  \}  \) is a closed set.
\end{proposition}
\begin{proof}
We proceed by induction on the structure of \( D \).
    \begin{itemize}
    \item $D \equiv F $: It immediately follows from Lemma~\ref{lem:State_closed}.
      \item  $D\equiv \oplus_{i \in I} p_i F_i $: Since
    $[\![\oplus_{i \in I} p_i F_i ]\!] = \{\sum_{i \in I} p_i \mu_i \, |\, \mu_i \in [\![F_i]\!]\} $, which is a linear combination of $\{[\![F_i]\!]\}_{i \in I}$ with weights $\{p_i\}_{i \in I}$.  
    By Lemma~\ref{lem:State_closed}, $[\![F_i]\!]$ is closed for any $ i$. It follows from lemma~\ref{combin} that $[\![\oplus_{i \in I} p_i F_i ]\!] $ is also closed.

    \item $D\equiv\oplus_{i\in I}F_i$: Note that $[\![\oplus_{i\in I}F_i]\!] = \{\mu \mid \mu=\sum_{i\in I}p_i\mu_i \} $  for some $p_i$ such that $\sum_{i\in I}p_i=1$ and $\mu_i$ such that  $\mu_i\in [\![F_i]\!]$. By Lemma~\ref{lem:State_closed}, the set $[\![F_i]\!]$ is closed for each $i\in I$. It follows from Lemma~\ref{Con} that $[\![\oplus_{i\in I}F_i]\!]$ is a closed set. \qedhere
    \end{itemize}
\end{proof}

\section{Proof outlines for functional correctness of programs}

\label{appB}

\begin{longtable}{l}
\caption{Proof outline for the loop body of the HHL algorithm}\label{t:body1} \\
  $\Key{\{\  v=0 \ \}}$ \\
  $\Key{\{\ \true\ \} \qquad\mbox{by rule [PT]}}$\\
  $\Key{\{\ \true \odot \true \odot \true\ \} \qquad\mbox{by rule [OdotE]}}$\\
  $\InOut \Key{\{\  \mathbf{true}\ \}} \quad  \overline{p}:=\ket{0}^{\otimes n}; \quad \Key{\{\  \ket{0}^{\otimes n}_{\overline{p}} \ \} \qquad\mbox{by rule [QInit]}}$\\
  $\InOut \Key{\{\  \mathbf{true}\ \}} \quad  \overline{q}:=\ket{0}^{\otimes m}; \quad \Key{\{\  \ket{0}^{\otimes m}_{\overline{q}} \ \} \qquad\mbox{by rule [QInit]}}$\\
  $\InOut \Key{\{\  \mathbf{true}\ \}} \quad r:=\ket{0}; \quad \Key{\{\  \ket{0}_r \ \} \qquad\mbox{by rule [QInit]}}$\\
  $\InOut \Key{\{\ \ket{0}^{\otimes m}_{\overline{q}}\ \}} \quad  U_b[\overline{q}]; \quad \Key{\{\  \ket{b}_{\overline{q}} \ \} \qquad\mbox{by rule [QUnit]}}$\\
  $\InOut \Key{\{\ \ket{0}^{\otimes n}_{\overline{p}}\ \}} \quad  H^{\otimes n}[\overline{p}]; \quad \Key{\{\ \frac{1}{\sqrt{N}}\sum_{z=0}^{N-1}|z\>_{\overline{p}}\ \} \qquad\mbox{by rule [QUnit]}}$\\
  $\Key{\{\ (\frac{1}{\sqrt{N}}\sum_{z=0}^{N-1}|z\>_{\overline{p}})(\sum_j b_j|j\>_{\overline{q}})|0\>_r\ \} \qquad\mbox{by rule [QFrame]}} $ \\
  $\In \Key{\{\ \sum_j b_j\frac{1}{\sqrt{N}}\sum_{z=0}^{N-1}|z\>_{\overline{p}}|j\>_{\overline{q}}\ \}}$ \\
  $ U_f[\overline{p}\overline{q}]; $\\
  $ \Out \Key{\{\ \sum_j b_j\frac{1}{\sqrt{N}}\sum_{z=0}^{N-1} e^{2\pi i\phi_j z} |z\>_{\overline{p}}|j\>_{\overline{q}}\ \} \qquad\mbox{by rule [QUnit]}}$\\
  $ \mathrm{QFT}^{-1}[\overline{p}];$ \\
  $\Key{\{\ \sum_j b_j |\phi_j\cdot N\>_{\overline{p}}|j\>_{\overline{q}}|0\>_r\ \} \qquad\mbox{by rule [QUnit]}}$\\
  $ \Key{\{\ \sum_j b_j |\phi_j\cdot N\>_{\overline{p}}|0\>_r|j\>_{\overline{q}}\ \} \qquad\mbox{by rule [ReArr]}}$\\
  $ U_c[\overline{p}r]; $\\
  $ \Key{\{\ \sum_{j}  \mathit{b}_{j} \ket{\widetilde{\phi_{j}}}_{\overline{p}}  (\sqrt{1-\frac{C^2}{\widetilde{\phi_{j}}^2}}\ket{0}+\frac{C}{\widetilde{\phi_{j}}}\ket{1})_{r} \ket{j}_{\overline{q}}\ \} \qquad\mbox{by rule [QUnit]}}$\\
   $ \Key{\{\ \sum_{j}  \mathit{b}_{j} \ket{\widetilde{\phi_{j}}}_{\overline{p}}\ket{j}_{\overline{q}}  (\sqrt{1-\frac{C^2}{\widetilde{\phi_{j}}^2}}\ket{0}+\frac{C}{\widetilde{\phi_{j}}}\ket{1})_{r} \ \} \qquad\mbox{by rule [ReArr]}}$\\
$\mathrm{QFT}[\overline{p}];$\\
  $ \Key{\{\ \sum_{j=1}^{N}  \mathit{b}_{j} (  \dfrac{1}{\sqrt{N}} \sum_{z=0}^{N-1} e^{2 \pi i \phi_{j} z} \ket{z}_{\overline{p}})  \ket{j}_{\overline{q}}(\sqrt{1-\frac{C^2}{\widetilde{\phi_{j}}^2}}\ket{0}+\frac{C}{\widetilde{\phi_{j}}}\ket{1})_{r} )\ \} \qquad\mbox{by rule [QUnit]}}$\\
$ U_{f}^{\dagger}[\overline{p}\overline{q}];$\\
$ \Key{\{\ \sum_{j} \mathit{b}_{j} \dfrac{1}{\sqrt{N}} \sum_{z=0}^{N-1}  \ket{z}_{\overline{p}} \ket{j}_{\overline{q}} (\sqrt{1-\frac{C^2}{\widetilde{\phi_{j}}^2}}\ket{0}+\frac{C}{\widetilde{\phi_{j}}}\ket{1})_{r}   \ \}\qquad\mbox{by rule [QUnit]}}$\\
$ H^{\otimes n}[\overline{p}]$;\\
  $ \Key{\{\  \sum_{j} \mathit{b}_{j}  \ket{0}^{\otimes n}_{\overline{p}}\ket{j}_{\overline{q}} (\sqrt{1-\frac{C^2}{\widetilde{\phi_{j}}^2}}\ket{0}+\frac{C}{\widetilde{\phi_{j}}}\ket{1})_{r}  \ \} \qquad\mbox{by rule [QUnit]}}$\\
  $ \Key{\{\  \sum_{j} \mathit{b}_{j}  \ket{0}^{\otimes n}_{\overline{p}}\ket{j}_{\overline{q}} (\sqrt{1-\frac{C^2}{\widetilde{\phi_{j}}^2}}\ket{0}+\frac{C}{\widetilde{\phi_{j}}}\ket{1})_{r} \wedge (v=0)[0/v] \wedge (v=1)[1/v] \ \} \qquad\mbox{by rule [Conseq]}}$\\
  $v:=M[r];$\\
  $\Key{\{\ \sum_j|b_j|^2(1-\frac{C^2}{\widetilde{\phi_{j}}^2}) \cdot (\sum_{j} \mathit{b}_{j} \sqrt{1-\frac{C^2}{\widetilde{\phi_{j}}^2}} \ket{0}^{\otimes n}_{\overline{p}}\ket{j}_{\overline{q}}|0\>_r/\sqrt{\sum_j|b_j|^2(1-\frac{C^2}{\widetilde{\phi_{j}}^2})}) \wedge (v=0) }$ \\
  $\Key{\oplus \sum_j|\frac{b_j C}{\widetilde{\phi_j}}|^2 \cdot (\sum_{j} \mathit{b}_{j} \frac{C}{\widetilde{\phi_j}} \ket{0}^{\otimes n}_{\overline{p}}\ket{j}_{\overline{q}}|1\>_r/\sqrt{\sum_j|\frac{b_j C}{\widetilde{\phi_j}}|^2}) \wedge (v=1) \ \}\qquad\mbox{by rule [QMeas]}}$\\
   $\Key{\{\  (\sum_{j} \mathit{b}_{j}\sqrt{1-\frac{C^2}{\widetilde{\phi_{j}}^2}}  \ket{0}^{\otimes n}_{\overline{p}}\ket{j}_{\overline{q}}|0\>_r/\sqrt{\sum_j|b_j|^2(1-\frac{C^2}{\widetilde{\phi_{j}}^2})}) \wedge (v=0)} $ \\
  $\Key{\oplus  (\sum_{j} \frac{b_j C}{\widetilde{\phi_j}}  \ket{0}^{\otimes n}_{\overline{p}}\ket{j}_{\overline{q}}|1\>_r/\sqrt{\sum_j|\frac{b_j C}{\widetilde{\phi_j}}|^2}) \wedge (v=1) \ \}\qquad\mbox{by rule [Oplus]}}$\\
  $\Key{\{\  (v=0) \oplus  ((\sum_{j} \frac{b_j}{\lambda_j}  \ket{0}^{\otimes n}_{\overline{p}}\ket{j}_{\overline{q}}|1\>_r/\sqrt{\sum_j|\frac{b_j}{\lambda_j}|^2}) \wedge (v=1)) \ \} \qquad\mbox{by rule [Conseq]}}$\\
  $\Key{\{\  (v=0) \oplus  (( \ket{0}^{\otimes n}_{\overline{p}}\ket{x}_{\overline{q}}|1\>_r) \wedge (v=1)) \ \} \qquad\mbox{by (\ref{e:x}) and rule [Conseq]}}$\\
  \\
\end{longtable}

\begin{longtable}[t]{ll}
  \caption{Proof outline of the loop body of the \textbf{OF} program}\label{t:body2}\\
 & $\Key{\{\ z < r \wedge b\not=1  \ \}}$\\
 & $\Key{\{\ z < r \wedge b\not=1 \odot \true \odot\true \ \}\qquad\mbox{by rule [OdotE]}}$\\
 & $\InOut \Key{\{\  \true \ \}}\quad \overline{q}:=\ket{0}^{\otimes t};\quad \Key{\{\ \ket{0}_{\overline{q}}^{\otimes t} \ \}\qquad\mbox{by rule [QInit]}}$\\
 & $\InOut \Key{\{\  \true \ \}}\quad \overline{p}:=\ket{0}^{\otimes L};\quad \Key{\{\ \ket{0}_{\overline{p}}^{\otimes L} \ \}\qquad\mbox{by rule [QInit]}}$\\
 & $\InOut \Key{\{\  \ket{0}_{\overline{q}}^{\otimes t} \ \}}\quad H^{\otimes t}[\overline{q}];\quad \Key{\{\ \ket{+}_{\overline{q}}^{\otimes L} \ \}\qquad\mbox{by rule [QUnit]}}$\\
 & $\InOut \Key{\{\  \ket{0}_{\overline{p}}^{\otimes L} \ \}}\quad  U_{+}[\overline{p}];\quad \Key{\{\ \ket{1}_{\overline{p}} \ \}\qquad\mbox{by rule [QUnit]}}$\\
  & $ \Key{\{\ \frac{1}{\sqrt{r2^t}}\sum_{s=0}^{r-1}\sum_{j=0}^{2^t-1}\ket{j}_{\overline{q}}\ket{u_s}_{\overline{p}} \ \} \qquad\mbox{by rule [QFrame]}} $\\
  & $CU[\overline{q}\overline{p}];$\\
 &  $ \Key{\{\ \frac{1}{\sqrt{r2^t}}\sum_{s=0}^{r-1}\sum_{j=0}^{2^t-1}e^{2\pi ijs/r}\ket{j}_{\overline{q}}\ket{u_s}_{\overline{p}} \ \}\qquad\mbox{by rule [QUnit]}}$\\
 & $ \Key{\{
	\frac{1}{\sqrt{r}}\sum_{s=0}^{r-1}(\frac{1}{\sqrt{2^t}}\sum_{j=0}^{2^t-1} e^{ \frac{2\pi i}{2^t} j \frac{s}{r} 2^t}\ket{j}_{\overline{q}})  \ket{u_{s}}_{\overline{p}}\}\qquad\mbox{by rule [Conseq]}}$\\
& $\mathrm{QFT^{-1}}[\overline{q}];$\\
 & $\Key{\{\frac{1}{\sqrt{r}} \sum_{s=0}^{r-1} \ket{\frac{s}{r}2^t}_{\overline{q}} \ket{u_{s}}_{\overline{p}}\}\qquad\mbox{by rule [QUnit]}}$\\
  & $\Key{\{\frac{1}{\sqrt{r}} \sum_{s=0}^{r-1} \ket{\frac{s}{r}2^t}_{\overline{q}} \ket{u_{s}}_{\overline{p}}\wedge \wedge_{s=0}^{r-1}(z'=\frac{s}{r}2^t)[\frac{s}{r}2^t/z']\}\qquad\mbox{by rule [Conseq]}}$\\
  & $z':=M[\overline{q}];$ \\
  & $\Key{\{ \oplus_{s} \frac{1}{r}\cdot (
	\ket{\frac{s}{r}2^t}_{\overline{q}}  \ket{u_{s}}_{\overline{p}} \wedge z'=\frac{s}{r}2^t)\}\qquad\mbox{by rule [QMeas]}}$\\
 &  $\Key{\{ \oplus_{s} 
    \ket{\frac{s}{r}2^t}_{\overline{q}}  \ket{u_{s}}_{\overline{p}} \wedge z'=\frac{s}{r}2^t\}\qquad\mbox{by rule [Oplus]}}$\\
 &  $\Key{\{ \oplus_{s}  z'=\frac{s}{r}2^t\}\qquad\mbox{by rule [Conseq]}}$\\
  %     $\Key{\{ \oplus_{s} (gcd(s,r)=1\vee gcd(s,r)\not=1)  z'=\frac{s}{r}2^t\}\qquad\mbox{by rule [Conseq]}}$\\
 & $\Key{\{ \oplus_{s}  ((z=r \oplus z < r)[f(\frac{z'}{2^t})/z])\}\qquad\mbox{by rule [Conseq]}}$  \\
 & $z:=f(\frac{z'}{2^t})$;\\
  % $\Key{\{ \oplus_{s}  (z'=\frac{s}{r}2^t \wedge (z=r \oplus z\not=r))\}\qquad\mbox{by rule [Assgn]}}$\\
 & $\Key{\{ z=r \oplus z < r\}\qquad\mbox{by rule [Assgn]} }$\\
  & $\Key{\{ ((z=r\wedge b=1) \oplus (z < r\wedge b\not=1)) [(x^z\ \mathrm{mod}\ N)/b]\}\qquad\mbox{by rule [Conseq]}}$\\
  & $b:=x^{z}\ \mathrm{mod} \ N;$\\
 & $\Key{\{ (z=r\wedge b=1) \oplus (z < r\wedge b\not=1)\}\qquad\mbox{by rule [Assgn]}}$ 
\end{longtable}

	\begin{longtable}[t]{ll}
    \caption{Proof outline for Shor's algorithm} \label{t:shor}\\
	& $\Key{\{ {\rm cmp}(N)\}}$\\
	1: & $\mathbf{if}\ 2 \mid N \ \mathbf{then}$ \\
	&\qquad $\Key{\{{\rm cmp}(N) \wedge N \  \mathrm{mod} \  2=0 \}}$\\
 &\qquad $\Key{\{ {\rm cmp}(N) \wedge N \ \mathrm{mod} \  2=0 \wedge (y=2)[2/y]\}\quad\mbox{by rule [Conseq]} }$ \\
  &\qquad $\In \Key{\{{ (y=2)[2/y]\} }} $ \\
	2: & \qquad $y:=2$; \\
	&\qquad $\Out\Key{\{y=2\}\quad\mbox{by rule [Assgn]}}$\\
 &\qquad $\Key{\{ {\rm cmp}(N) \wedge N \ \mathrm{mod} \  2=0 \wedge y=2 \}\quad\mbox{by rule [Qframe]}}$\\
	&\qquad $\Key{ \{ y|N \wedge y\neq 1\wedge y\neq N \}\quad\mbox{by rule [Conseq]}}$\\
	3: & $\mathbf{else}$ \\
	&\qquad $\Key{\{ {\rm cmp}(N) \wedge N \ \mathrm{mod} \ 2 \neq 0  \}}$\\
 %   &\qquad $\Key{\{ {\rm cmp}(N) \wedge N \ \mathrm{mod} \ 2 \neq 0 \wedge ( x= \mathrm{random}(2,N-1) )[\mathrm{random}(2,N-1)/x]
 % \}\quad\mbox{by rule [Conseq]}}$\\
        &\qquad $\Key{\{ (2\leq x\leq N-1)[\mathrm{random}(2,N-1)/x]
 \}\quad\mbox{by rule [Conseq]}}$\\
 %  &\qquad $\In \Key{\{ (2\leq x\leq N-1)[\mathrm{random}(2,N-1)/x] 
 % \}}$\\
	4:& \qquad $x:=\mathrm{random}(2,N-1)$; \\
        &\qquad $ \Key{\{ (2\leq x\leq N-1) \}\quad\mbox{by rule [Assgn]}}$\\
           & \qquad  $\Key{\{ (2\leq x\leq N-1) \wedge (y= \mathrm{gcd}(x,N))[\mathrm{gcd}(x,N)/y] \}\quad\mbox{by rule [Conseq]}}$\\
       & \qquad  $\In \Key{\{ (y= \mathrm{gcd}(x,N))[\mathrm{gcd}(x,N)/y] \}}$\\
    5:& \qquad $y:=\gcd(x,N)$; \\
    & \qquad  $\Out \Key{\{ (y= \mathrm{gcd}(x,N)) \}\quad\mbox{by rule [Assgn]}}$\\
     & \qquad  $ \Key{\{(2\leq x\leq N-1) \wedge (y= \mathrm{gcd}(x,N))\}\quad\mbox{by rule [Qframe]}}$\\
     &\qquad \textcolor{red}{$\{ (2\leq x\leq N-1) \wedge ((y= \gcd (x, N) \wedge y\neq N) $ } \\
       &\quad  \qquad  \qquad \qquad \qquad \quad \textcolor{red}{$  \vee (y= \gcd (x^{r/2}-1, N) \wedge y\neq N \wedge y \neq 1)$ } \\
     & \quad \qquad \qquad \qquad  \qquad \quad  \textcolor{red}{$ \vee (y= \gcd (x^{r/2}+1, N) \wedge y\neq N \wedge y \neq 1))$ \}  \quad \mbox{by rule [Conseq]} }\\
	6:& \qquad $\mathbf{while} \ y=1 \ \mathbf{do}$\\
  &\qquad\qquad  \Key{\{  $(2\leq x\leq N-1) \wedge \gcd (x, N) =1$\} 
\quad \mbox{by rule [Conseq]}  } \\
     % &\qquad\qquad $ \Key{\{ y|N \wedge y\neq N \wedge y =1 \wedge z=\mathbf{ord}(x,N)[OF(x,N)/z]\} \quad\mbox{by rule [Conseq]}}$\\
    7:& \qquad \qquad $ OF(x,N)$; \\
     &\qquad\qquad $ \Key{\{  (2\leq x\leq N-1) \wedge  z=\mathbf{ord}(x,N)\} }$\\
    8:& \qquad \qquad $\mathbf{if}\ 2\mid z \mbox{ and } x^{z/2}\not\equiv -1\ (\mathrm{mod}\ N) \ \mathbf{then}$ \\
    &\qquad\qquad \qquad $\Key{\{ (2\leq x\leq N-1) \wedge  z=\mathbf{ord}(x,N) \wedge 2\mid z \wedge x^{z/2}\not\equiv_{N} -1  \} }$\\
    % &\qquad \qquad \qquad $\Key{\{ (2\leq x\leq N-1) \wedge  z=\mathbf{ord}(x,N) (x^{z/2})^2 \ \mathrm{mod} \ N=1 \wedge x^{z/2}\neq_{N}-1 \wedge x^{r/2}\neq_{N} 1\}\quad\mbox{by rule [Conseq]}}$\\
     &\qquad \qquad \qquad $\Key{\{ (2\leq x\leq N-1) \wedge  z=\mathbf{ord}(x,N) }$\\
      &\qquad \qquad \qquad \quad $\Key{\wedge ((\mathrm{gcd}(x^{z/2}-1,N) \neq 1 \wedge \mathrm{gcd}(x^{z/2}-1,N) \neq N)}$\\
      &\qquad \qquad \qquad \qquad  $\Key{ \vee (\mathrm{gcd}(x^{z/2}+1,N)\neq 1 \wedge \mathrm{gcd}(x^{z/2}+1,N) \neq N)) \}\quad\mbox{by rule [Conseq]}}$\\
 %     & \qquad \qquad \qquad $\In \Key{\{ (y'|N \wedge y'|(x^{z/2}-1))[\mathrm{gcd}(x^{z/2}-1,N)/y'] \}}$\\
	% 9:& \qquad \qquad \qquad $y':=\mathrm{gcd}(x^{z/2}-1,N)$; \\
 %   & \qquad \qquad \qquad $\Out \Key{\{ (y'|N \wedge y'|(x^{z/2}-1)) \} \quad \mbox{by rule [Assgn]}}$\\
	9:& \qquad \qquad \qquad $\mathbf{if}\ 1<(\mathrm{gcd}(x^{z/2}-1,N) <N \ \mathbf{then}$ \\
    &\qquad \qquad \qquad $\Key{\{ (2\leq x\leq N-1) \wedge  z=\mathbf{ord}(x,N) \wedge (\mathrm{gcd}(x^{z/2}-1,N) \neq 1 \wedge \mathrm{gcd}(x^{z/2}-1,N) \neq N)  \} }$\\
      % &\qquad \qquad \qquad \quad $\Key{\wedge ((\mathrm{gcd}(x^{z/2}-1,N) \neq 1 \wedge \mathrm{gcd}(x^{z/2}-1,N) \neq N)}$\\
       &\qquad \qquad \qquad $\Key{\{ (2\leq x\leq N-1)} $ \\ 
       & \qquad \qquad \qquad $\Key{\wedge (y= \gcd (x^{r/2}-1, N) \wedge y\neq N \wedge y \neq 1) [\mathrm{gcd}(x^{z/2}-1,N)/y]\} \quad\mbox{by rule [Conseq]}  }$ \\
          &\qquad \qquad \qquad $\In \Key{\{ (y= \gcd (x^{r/2}-1, N) \wedge y\neq N \wedge y \neq 1) [\mathrm{gcd}(x^{z/2}-1,N)/y]\} }$\\
10:& \qquad \qquad \qquad $y :=\mathrm{gcd}(x^{z/2}-1,N)$; \\
 &\qquad \qquad \qquad  $\Out \Key{ \{ (y= \gcd (x^{r/2}-1, N) \wedge y\neq N \wedge y \neq 1)\} \quad\mbox{by rule [Assgn]}  }$\\
      &\qquad \qquad \qquad  $\Key{ \{ (2\leq x\leq N-1) \wedge (y= \gcd (x^{r/2}-1, N) \wedge y\neq N \wedge y \neq 1)\} \quad\mbox{by rule [Qframe]} }$\\
 &\qquad \qquad \qquad \Key{$\{ (2\leq x\leq N-1) \wedge ((y= \gcd (x, N) \wedge y\neq N) $ } \\
       &\qquad  \qquad  \qquad \qquad \qquad \quad \Key{$  \vee (y= \gcd (x^{r/2}-1, N) \wedge y\neq N \wedge y \neq 1)$ } \\
     & \qquad \qquad \qquad \qquad  \qquad \quad \Key{$ \vee (y= \gcd (x^{r/2}+1, N) \wedge y\neq N \wedge y \neq 1))$ \}  \quad \mbox{by rule [Conseq]} }\\
    11:& \qquad \qquad \qquad $\mathbf{else}$ \\
     &\qquad \qquad \qquad $\Key{\{ (2\leq x\leq N-1) \wedge  z=\mathbf{ord}(x,N) \wedge (\mathrm{gcd}(x^{z/2}+1,N) \neq 1 \wedge \mathrm{gcd}(x^{z/2}+1,N) \neq N) \}  }$\\
  &\qquad \qquad \qquad $\Key{\{ (2\leq x\leq N-1) }$\\
  & \qquad \qquad \qquad 
$ \Key{\wedge (y= \gcd (x^{r/2}+1, N) \wedge y\neq N \wedge y \neq 1) [\mathrm{gcd}(x^{z/2}+1,N)/y]  \}\quad\mbox{by rule [Conseq]} }$\\
          &\qquad \qquad \qquad $\In \Key{\{ (y= \gcd (x^{r/2}+1, N) \wedge y\neq N \wedge y \neq 1) [\mathrm{gcd}(x^{z/2}+1,N)/y] \} }$\\
    12:& \qquad \qquad \qquad \qquad $y:=\mathrm{gcd}(x^{z/2}+1,N)$; \\
  &\qquad \qquad \qquad  $\Out \Key{ \{ (y= \gcd (x^{r/2}+1, N) \wedge y\neq N \wedge y \neq 1)\} \quad\mbox{by rule [Assgn]}  }$\\
      &\qquad \qquad \qquad  $\Key{ \{ (2\leq x\leq N-1) \wedge (y= \gcd (x^{r/2}+1, N) \wedge y\neq N \wedge y \neq 1)\} \quad\mbox{by rule [Qframe]} }$\\
 &\qquad \qquad \qquad \Key{$\{ (2\leq x\leq N-1) \wedge ((y= \gcd (x, N) \wedge y\neq N) $ } \\
       &\qquad  \qquad  \qquad \qquad \qquad \quad \Key{$  \vee (y= \gcd (x^{r/2}-1, N) \wedge y\neq N \wedge y \neq 1)$ } \\
     & \qquad \qquad \qquad \qquad  \qquad \quad \Key{$ \vee (y= \gcd (x^{r/2}+1, N) \wedge y\neq N \wedge y \neq 1))$ \}  \quad \mbox{by rule [Conseq]} }\\
    13:& \qquad \qquad \qquad $\mathbf{fi}$ \\
     &\qquad \qquad \qquad \Key{$\{ (2\leq x\leq N-1) \wedge ((y= \gcd (x, N) \wedge y\neq N) $ } \\
       &\qquad  \qquad  \qquad \qquad \qquad \quad \Key{$  \vee (y= \gcd (x^{r/2}-1, N) \wedge y\neq N \wedge y \neq 1)$ } \\
     & \qquad \qquad \qquad \qquad  \qquad \quad \Key{$ \vee (y= \gcd (x^{r/2}+1, N) \wedge y\neq N \wedge y \neq 1))$ \}  \quad \mbox{by rule [Cond]} }\\
     % & \qquad \qquad \qquad  $ \Key{\{ (y|N \wedge y \neq 1 \wedge y \neq N) \} \quad \mbox{by rule [Cond]}}$\\
    14:& \qquad \qquad $\mathbf{else}$ \\
       % & \qquad \qquad \qquad  $ \Key{\{ z=\mathbf{ord}(x,N) \wedge \neg(2\mid z \wedge x^{z/2}\not\equiv_{N} -1) \}} $\\
       &\qquad \qquad \qquad $\Key{\{ (2\leq x\leq N-1)[\mathrm{random}(2,N-1)/x] \}\quad\mbox{by rule [Conseq]}}$\\
    15:& \qquad \qquad \qquad $x:=\mathrm{random}(2,N-1)$; \\
    &\qquad \qquad \qquad $\Key{\{ (2\leq x\leq N-1) \}\quad\mbox{by rule [Assgn]}}$\\
     & \qquad  \qquad \qquad $\Key{\{ (2\leq x\leq N-1) \wedge (y= \mathrm{gcd}(x,N))[\mathrm{gcd}(x,N)/y] \}\quad\mbox{by rule [Conseq]}}$\\
       & \qquad \qquad \qquad $\In \Key{\{ (y= \mathrm{gcd}(x,N))[\mathrm{gcd}(x,N)/y] \}}$\\
    16:& \qquad \qquad \qquad $y:=\gcd(x,N)$; \\
    & \qquad  \qquad \qquad $\Out \Key{\{ (y= \mathrm{gcd}(x,N)) \}\quad\mbox{by rule [Assgn]}}$\\
     & \qquad  \qquad \qquad $ \Key{\{(2\leq x\leq N-1) \wedge (y= \mathrm{gcd}(x,N))\}\quad\mbox{by rule [Qframe]}}$\\
  &\qquad \qquad  \qquad \Key{$\{ (2\leq x\leq N-1) \wedge ((y= \gcd (x, N) \wedge y\neq N) $ } \\
       &\qquad   \qquad \qquad \qquad \quad \Key{$  \vee (y= \gcd (x^{r/2}-1, N) \wedge y\neq N \wedge y \neq 1)$ } \\
     & \qquad  \qquad \qquad  \qquad \quad \Key{$ \vee (y= \gcd (x^{r/2}+1, N) \wedge y\neq N \wedge y \neq 1))$ \}  \quad \mbox{by rule [Conseq]} }\\
    17:& \qquad \qquad $\mathbf{fi}$ \\
    &\qquad \qquad  \textcolor{red}{$\{ (2\leq x\leq N-1) \wedge ((y= \gcd (x, N) \wedge y\neq N) $ } \\
       &\qquad  \qquad  \qquad \qquad \qquad \quad \textcolor{red}{$  \vee (y= \gcd (x^{r/2}-1, N) \wedge y\neq N \wedge y \neq 1)$ } \\
     & \qquad \qquad \qquad \qquad  \qquad \quad  \textcolor{red}{$ \vee (y= \gcd (x^{r/2}+1, N) \wedge y\neq N \wedge y \neq 1))$ \}  \quad \mbox{by rule [Cond]} }\\
    18:& \qquad $\mathbf{od}$ \\
    &\qquad $\Key{\{ y |N \wedge y \neq N  \wedge y \neq 1\} \quad \mbox{by rule [While] and [Conseq]} }$\\
    19:& $\mathbf{fi}$ \\
    & $\Key{\{ y |N \wedge y \neq N  \wedge y \neq 1\} \quad \mbox{by rule [Cond]} }$\\
    & \\
\end{longtable}

\end{document}